\documentclass[5p,10pt,fleqn]{elsarticle}

\usepackage{lineno}

\usepackage[utf8]{inputenc}
\usepackage[english]{babel}
\usepackage{amssymb}
\usepackage{amsfonts}
\usepackage{amsmath}
\usepackage{txfonts}
\usepackage{graphicx}
\usepackage{ae,aecompl}
\usepackage{fancyhdr}
\usepackage{multicol}
\usepackage{layout}
\usepackage{graphicx}
\usepackage{multirow}
\usepackage{times}
\usepackage{textcomp}
\usepackage{cprotect}
\usepackage[dvipsnames]{xcolor}
\usepackage{enumitem}
\usepackage{xcolor}

\usepackage[outdir=./figures/]{epstopdf}
\usepackage[export]{adjustbox}

\usepackage{comment}

\usepackage{hyperref}
\hypersetup{%
    ,urlcolor=blue
    ,citecolor=blue
    ,linkcolor=blue
    }
    
\graphicspath{{./figures/}}

\biboptions{sort&compress}

\newif\ifAMStwofonts
\AMStwofontstrue

\begin{document}

\begin{frontmatter}

\title{Effects of dynamical friction on perturbations for evolving dark energy}

\author[1,11,111]{Francesco Pace\corref{cor1}}
\ead{francesco.pace@unito.it}

\affiliation[1]{organization={Dipartimento di Fisica, Universit\`a degli Studi di Torino},
            addressline={Via P. Giuria 1}, 
            postcode={I-10125}, 
            city={Torino},
            country={Italy}}

\affiliation[11]{organization={INFN-Sezione di Torino},
            addressline={Via P. Giuria 1}, 
            postcode={I-10125}, 
            city={Torino},
            country={Italy}}

\affiliation[111]{organization={INAF-Istituto Nazionale di Astrofisica, Osservatorio Astrofisico di Torino},
            addressline={strada Osservatorio 20}, 
            postcode={10025}, 
            city={Pino Torinese},
            country={Italy}}

\author[2,22,222,2222]{Orlando Luongo}
\ead{orlando.luongo@unicam.it}

\affiliation[2]{organization={Department of Mathematics and Physics, SUNY Polytechnic Institute},
addressline={100 Seymour Rd},
postcode={13502},
city={Utica},
state={New York},
country={USA}}

\affiliation[22]{organization={Divisione di Fisica, Universit\`a degli studi di Camerino},
addressline={Via Madonna delle Carceri 9B},
postcode={62032},
city={Camerino},
country={Italy}}

\affiliation[222]{organization={INFN-Sezione di Perugia},
addressline={Via Alessandro Pascoli 23c},
postcode={06123},
city={Perugia},
country={Italy}}

\affiliation[2222]{organization={Al-Farabi Kazakh National University},
addressline={Al-Farabi Avenue 71},
postcode={050040},
city={Almaty},
country={Kazakhstan}}

\author[3,33,333]{Antonino Del Popolo}
\ead{antonino.delpopolo@unict.it}

\affiliation[3]{organization={Dipartimento di Fisica e Astronomia, University of Catania},
addressline={Viale Andrea Doria 6},
postcode={95125},
city={Catania},
country={Italy}}

\affiliation[33]{organization={INFN-Sezione di Catania},
addressline={Via Santa Sofia 64},
postcode={95123},
city={Catania},
country={Italy}}

\affiliation[333]{organization={Institute of Astronomy, Russian Academy of Sciences},
postcode={119017},
addressline={Pyatnitskaya str., 48},
city={Moscow},
country={Russia}}

\cortext[cor1]{Corresponding author}

\begin{abstract}
We explore the impact of dynamical friction on scales where the linear growth factor and the spherical collapse model can be applied and show its influence on the evolution of perturbations. In particular, considering smooth and clustering dark energy models, we describe the role played by friction by selecting two main hierarchical models, \textit{i.e.}, the first where the friction term is proportional to the Hubble rate, whereas the second where friction is induced by the dark energy pressure. The second approach generalises the first and translates the idea that pressure is a general relativistic effect, motivating why friction might arise once barotropic dark energy fluids are considered. The corresponding effects of friction are investigated at the level of linear and nonlinear perturbations, using the formalism of the spherical collapse model. Whilst dynamical friction has very small effects and thus it cannot be excluded \emph{a priori}, dissipative pressure friction leads to a substantial slow down in the evolution of perturbations. This can be inferred particularly from the halo mass function, for which we also employ corrections due to dark energy clustering. To this end, in order to discern detectable deviations from the standard cosmological model, we thus highlight where dissipation effects might play a significant role at large scales. 
\end{abstract}


\end{frontmatter}

\section{Introduction}

Ever since the discovery of cosmic acceleration \cite{Riess1998,Perlmutter1998}, several probes undoubtedly showed that the universe expansion rate is currently speeding up, through something unknown called dark energy, or in the simplest case by vacuum energy \cite{Martin2012}. In order to disclose the dark energy nature, several theoretical models have been proposed throughout the last two decades \cite{Copeland2006}. 

In addition to the existence of low- and high-redshift tests \cite{Aviles2012}, alternatives can make use of morphological features of the observable universe, among which high-precision galaxy clustering and weak gravitational lensing measurements, to check how dark matter can cluster. The clustering process is intimately related to small perturbations induced by cosmic fluids in the universe, and, in fact, any dark energy model might successfully overcome the aforementioned low- and high-redshift tests, together with morphological probes.

In this respect, \emph{concordance background paradigm}, \textit{i.e.}, the best candidate to explain how the acceleration occurs, dubbed the $\Lambda$CDM model \cite{Bull:2015stt}, where $\Lambda$ represents the cosmological constant and CDM the cold dark matter component, unequivocally appears to be the best description of the Cosmic Microwave Background (CMB) and Large Scale Structure (LSS) surveys. Phrasing it differently, the concordance approach depends on the fewest number of free parameters possible \cite{Perivolaropoulos:2021jda}, suggesting that any dark energy model is \emph{statistically disfavoured} with respect to it \cite{Shafieloo:2018gin}\footnote{For the sake of clarity, the $\Lambda$CDM model requires six parameters, albeit at late-times, one single parameter, namely the mass density, approximates well the overall dynamics. Thus, to falsify the model, a compelling dark energy scenario might be described by one parameter only. The only class of dark energy models that prompts one parameter only, being different from $\Lambda$CDM is \emph{dark fluid}, see e.g. \cite{Arbey:2006it,Aviles:2011ak,Luongo:2018lgy,DAgostino:2022fcx}. Model-independent tests seem to certify the goodness of the standard background model at both low and intermediate levels \cite{Luongo:2020aqw}.}.

So, apparently, there is no reason to go further, as the $\Lambda$CDM model describes the universe so well at large scales. However, even though the cosmological constant $\Lambda$, within the context of general relativity, describes well the cosmic speed up, new dark sector physics seems to be required in order to heal current tensions \cite{DiValentino:2021izs,Douspis:2018xlj} and the associated caveats related to the $\Lambda$CDM model itself.

Consequently, a complete understanding of nonlinear structure formation becomes central to model the power spectrum and then to infer deviations from the concordance paradigm. In this respect, rich clusters can be used to constrain cosmological models, as they represent the largest virialized objects in the universe\footnote{Hence, they can be put in relation with the linear matter power spectrum, whereas their temperature can be used to infer the cluster mass.}.

Several alternatives to the standard concordance model have been investigated \cite{Copeland2006,Capozziello:2019cav}, with models spanning from extending general relativity, up to barotropic dynamical dark energy, passing through interacting dark energy models \citep{Sharma2022,Sharma2023}\footnote{Most of these approaches assume some form of energy transfer that affects both background and perturbations. These models often fail to fit the CMB data or are severely constrained.}.

Among all, the study of friction at cosmological scales and dynamical friction, namely the friction supposed to evolve according to the equation of state of dark energy, represent plausible features of cosmological perturbations not yet fully explored. Plausible effects induced by dissipation may compete against clustering during the overall process that induces the overdensity formation, suggesting that including friction is of crucial interest for understanding, in dark energy scenarios, how clustering is modified accordingly.

Motivated by the above points, we investigate here the effects resulting from dynamical friction at large scales. In this respect, we consider both smooth and clustering dark energy models, limiting the analysis, for the sake of simplicity, to models with a linearly evolving equation of state \cite{Chevallier2001,Linder2003}. We choose this class of models to ensure that the observations hold, as dark energy scenarios seem to favor a weakly evolving dark energy term, instead of more complicated functions \cite{Capozziello:2014zda,Gruber:2013wua,Capozziello:2015rda,Dunsby:2015ers,Aviles:2014rma,Capozziello:2020ctn,Aviles:2013nga,Aviles:2012ir,Luongo:2011zz,delaCruz-Dombriz:2016bqh,Capozziello:2018aba,Capozziello:2017ddd,Luongo:2013rba,Aviles:2016wel}\footnote{For these reasons, we exclude extended theories of gravity and we focus on general relativity only. Cosmic tests seem to agree with this assumption \cite{Calza:2019egu,Muccino:2020gqt}.}. In particular, to describe the role that friction plays, we have singled out two main models. In the first scenario, friction is proportional to the Hubble rate, whereas the second shows a more complicated correction that is mostly proportional to the dark energy pressure. The corresponding effects of friction are investigated at the level of linear and nonlinear perturbations, adopting the basic demands of spherical collapse theory. The evolution of perturbations is studied and the initial conditions properly inferred. We display our numerical bounds, showing that the main discrepancy occurs for the first hierarchy, at both the levels of linear and nonlinear perturbations. The consequences on the large-scale structures induced by the presence of friction are reported in detail, showing the main differences to standard frictionless cases.

Our numerical findings indicate that there is no reason to exclude \emph{a priori} the presence of friction in the cosmological realms. Even though the standard cosmological model does not take into account dissipation effects, we predict suitable bounds within the presence of dissipation, which can in principle be introduced. Since this property would significantly affect the overall clustering, we propose viable ranges to detect it.

The paper is organized as follows. In Sect.~\ref{sect:spcm} we present a detailed derivation of the equations of motion for the collapsing fluid and in Sect.~\ref{sect:friction} we discuss the models used in this work. In Sect.~\ref{sect:LinPert} we discuss the evolution of linear perturbations, while the study of nonlinear perturbations is presented in Sect.~\ref{sect:NonLinPert}. In Sect.~\ref{sect:MassFunction} we show the effects of dynamical friction on the halo mass function. We conclude our work in Sect.~\ref{sect:conclusions}.

\section{Extended spherical collapse model with friction}\label{sect:spcm}

In this section, we derive the expressions needed to study the evolution of linear and nonlinear perturbations of dark matter in a dark energy model, where the dark energy component is smooth and affects only the background evolution. We will then extend the formalism to include perturbations also for dark energy.

The spherical collapse model assumes that the perturbations are spherically symmetric and isolated \cite{Tomita1969,Gunn1972,Bernardeau1994,Lee2010}. A further assumption, which allows to greatly simplify the equations, is that perturbations are described by a top-hat profile, \textit{i.e.}, the profile does not change in space but it is constant over the size of the perturbation. The value of the density can therefore only change in amplitude, but the shape of the profile stays the same. This is a crude approximation, but it has been proven to be very effective in describing the nonlinear evolution of perturbations, showing a good agreement between the theoretical and numerical mass functions in $N$-body simulations for both dark energy \cite{Pace2010} and modified gravity models \cite{Pace2014}.

This basic picture has been improved in several ways, mainly relaxing the assumption of spherical symmetry and isolated perturbations. \cite{Ohta2003,Ohta2004} assumed triaxial (ellipsoidal) perturbations and took into account tidal fields; \cite{DelPopolo2013a,DelPopolo2013b,DelPopolo2013c,Pace2014b} considered the effects of shear and rotation (we will define them in a precise way later) assuming an expression derived by modeling the angular momentum of galaxies and clusters, while \cite{Reischke2016a,Pace2017,Reischke2018} proposed an approach based on the Zel'dovich approximation. Finally, \cite{Pace2019b,Pace2022} used deviations from sphericity to study the virialization of cosmic structures and \cite{Abramo2007,Abramo2008,Chang2018} considered perturbations also in the dark energy component.

Here, we build upon all these works by further extending the formalism to include the effects of dynamical friction in $\Lambda$CDM and dark energy models. Hence, we will derive the equations of motion for matter perturbations taking into account shear, rotation, and dynamical friction.

Dynamical friction manifests itself during the evolution of perturbations and affects their evolution, slowing their collapse. This means that cosmic structures require more time to form. As in Newtonian dynamics, friction affects the equations of motion and, being proportional to velocity, it modifies the Euler equation, which now reads
\begin{flalign}\label{eqn:Euler}
 \frac{\partial\mathbf{u}}{\partial t} + (2H+\eta)\mathbf{u} + \left(\mathbf{u}\cdot\nabla_{\mathbf{x}}\right)\mathbf{u} + \frac{1}{a^2}\nabla_{\mathbf{x}}\phi = 0\,. &&
\end{flalign}

In the previous equation, $H$ is the Hubble function, $\mathbf{u}$ represents the comoving peculiar velocity defined as $\mathbf{v} = aH\mathbf{x} + a\mathbf{u}$, with $a$ the scale factor, $\mathbf{x}$ the comoving position and $\mathbf{v}$ the fluid velocity. The perturbed potential is denoted by $\phi$. Finally, $\eta=\eta(t)$ is an arbitrary function of time that denotes the evolution of dynamical friction. As it is evident from Eq.~(\ref{eqn:Euler}), dynamical friction has dimensions of an inverse of time.

We note that this formulation, which later we will show to agree with the results derived based on the radius of the perturbation (upon some modifications we will discuss in detail), is very similar to that of models of dark scattering \citep{Carrilho2022}, but with an important difference. Whilst for dark scattering models the function $\eta$ is fixed by the theory, for models affected by friction this is a free function, whose specific form depends on the specific model chosen. As we shall discuss later, we shall consider two different functional forms, which are derived from different physical assumptions. Hence, despite the striking similarity to dark scattering models, here the phenomenology is much richer.

Continuity and Poisson equations are not affected by dynamical friction and their functional form is as usual given, respectively, by
\begin{subequations}
\begin{flalign}
 &\, \delta^{\prime} + (1+\delta)\tilde{\theta} = 0\,, && \\
 &\, \nabla^2\phi = \frac{3}{2}a^2H^2\Omega_{\rm m}(a)\delta\,, &&
\end{flalign}
\end{subequations}
with $\Omega_{\rm m}(a)$ the matter density parameter, $\delta=\delta\rho/\bar{\rho}$ the fluid density perturbation, $\tilde{\theta}=\theta/H$ with $\theta=\nabla\cdot\mathbf{u}$ being the divergence of the peculiar velocity and the prime represents the derivative with respect to $\ln{a}$.

We can combine the continuity, Euler and Poisson equations into a single equation describing the evolution of the perturbation $\delta$. Assuming as independent variable $\ln{a}$, the full nonlinear equation is
\begin{flalign}\label{eqn:delta_NL}
 \delta^{\prime\prime} + & \left(2 + \frac{\eta}{H} + \frac{H^{\prime}}{H}\right)\delta^{\prime} - \frac{4}{3}\frac{{\delta^{\prime}}^2}{1+\delta} - \frac{3}{2}\Omega_{\rm m}(a)\delta(1+\delta) - && \nonumber\\ 
 & \qquad 
 \left(\tilde{\sigma}^2-\tilde{\omega}^2\right)(1+\delta) = 0\,, &&
\end{flalign}
where $H^2\tilde{\sigma}^2 = \sigma^2 = \sigma_{ij}\sigma^{ij}$ and $H^2\tilde{\omega}^2 = \omega^2 = \omega_{ij}\omega^{ij}$ are the shear and rotation tensors, respectively. They are defined as
\begin{flalign}
 \sigma_{ij} = \frac{1}{2}(\partial_iu^j + \partial_ju^i) - \frac{1}{3}\theta\delta_{ij}\,, \quad
 \omega_{ij} = \frac{1}{2}(\partial_iu^j - \partial_ju^i)\,, &&
\end{flalign}
with $\delta_{ij}$ the Kronecker's delta symbol.

If we linearise Eq.~(\ref{eqn:delta_NL}), we obtain the modified equation describing the evolution of the growth factor:
\begin{flalign}\label{eqn:delta_L}
 \delta^{\prime\prime} + \left(2 + \frac{\eta}{H} + \frac{H^{\prime}}{H}\right)\delta^{\prime} - \frac{3}{2}\Omega_{\rm m}(a)\delta = 0\,, &&
\end{flalign}

Both the linear and nonlinear expressions assume the same form of those obtained for dark scattering models \cite{Baldi2017,Carrilho2021,Carrilho2022}. In dark scattering models, there is an interaction between the cold dark matter and the dark energy component, which gives rise to an exchange of energy between the two fluids. The equations of motion lead to an additional friction term which modifies the coefficient of $\delta^{\prime}$ and depends on the cross section and mass of the dark matter particles. However, we remark that the two models are physically distinct, and we mention dark-scattering models for completeness and for the similarity of the equations to be solved.

The formalism is easily extended to include the effects of clustering dark energy. To focus on this, let us now consider the following perturbation equations for the dark energy component \citep{Abramo2007,Abramo2008,Abramo2009a,Abramo2009b,Pace2017a}:
\begin{subequations}
 \begin{flalign}
  & \delta_\mathrm{de}^{\prime} + 3(s_{\rm eff}-w_\mathrm{de})\delta_\mathrm{de} + 
  \left[1+w_\mathrm{de}+\left(1+s_{\rm eff}\right)\delta_\mathrm{de}\right]\tilde{\theta} = \, 0\,,  \label{eqn:deltade}\\
  & \tilde{\theta}_\mathrm{de}^{\prime} + \left(2+\frac{H^{\prime}}{H}\right)\tilde{\theta}_\mathrm{de} + \frac{\tilde{\theta}_{\mathrm{de}}^2}{3} + 
  \frac{3}{2}\left[\Omega_\mathrm{m}(a)\delta_\mathrm{m}+(1+3s_{\rm eff})\Omega_\mathrm{de}(a)\delta_\mathrm{de}\right] = \, 0\,, && \label{eqn:theta}
 \end{flalign}
\end{subequations}
where the prime implies again the derivative with respect to $\ln{a}$. Clearly, $\tilde{\theta}_{\mathrm{de}}=\theta_{\mathrm{de}}/H$ and $w_\mathrm{de}$ is the dark energy equation of state in the background with the corresponding sound speed defined by $c_{\rm eff}^2=c^2s_{\rm eff}=\delta P_\mathrm{de}/\delta\rho_\mathrm{de}$, where the speed of light, $c$, appears for dimensional reasons. This quantity is zero for the standard $\Lambda$CDM model and for classes of dark fluids, while it is fixed to $s_{\rm eff}=1$ for quintessence. The different values that this can acquire have huge consequences in cosmology: a small sound speed implies structures to form at several scales, while high sound speeds correspond to stiff matter, suggesting perturbations to be strongly suppressed, \textit{i.e.}, involving dark energy that affects the background only. Consequently, it is licit to presume $s_{\rm eff}\ll 1$, guaranteeing that dark energy perturbations become relatively important in analogy to matter perturbations, albeit always subdominant \citep{Batista2013}. Note that, in general, $\theta_{\rm de} \neq \theta_{\rm m}$, as friction only affects (cold dark) matter.

\subsection{A comment on dynamical friction}

At this point, a deeper discussion of the results obtained is necessary. It is well known that an alternative way to discuss cosmological perturbations is based on the evolution of the physical radius of the perturbation $R$. In fact, many authors have already considered the effect of dynamical friction from this point of view. The evolution of the (effective) radius of the perturbation $R$, including angular momentum $L$ and dynamical friction $\eta$, in a $\Lambda$CDM model is given by \cite{Lahav1991,Bartlett1993,Peebles1993,AntonuccioDelogu1994,DelPopolo1998,DelPopolo1999}\footnote{In these references, only single parts of the full expression are investigated or reported.}
\begin{flalign}\label{eqn:Rddot_original}
 \frac{\mathrm{d}^2R}{\mathrm{d}t^2} = - \eta\frac{\mathrm{d}R}{\mathrm{d}t} - \frac{GM_{\rm m}}{R^2} + \frac{\Lambda c^2}{3}R + \frac{L^2}{M_{\rm m}R^2}\,, &&
\end{flalign}
where $t$ is the cosmic time, $c$ the speed of light, $G$ Newton's constant, $\Lambda$ the value of the cosmological constant and $M_{\rm m}$ the mass of the dark matter perturbation. This expression has been explicitly written down in \cite{DelPopolo2019} to derive a mass-temperature relation for clusters in $\Lambda$CDM and modified gravity models.

Following \cite{Pace2019b}, we can easily extend the previous expression to an arbitrary dark energy model described by an equation-of-state parameter $w_{\rm de}$ and express the angular momentum contribution in terms of the shear $\sigma$ and rotation $\omega$ terms
\begin{flalign}\label{eqn:eom_R}
 \frac{\mathrm{d}^2R}{\mathrm{d}t^2} = -\eta\frac{\mathrm{d}R}{\mathrm{d}t} - \frac{GM_{\rm m}}{R^2} - \frac{GM_{\rm bck,de}}{R^2}(1+3w_{\rm de}) -  \frac{\sigma^2-\omega^2}{3}R \,, &&
\end{flalign}
where we defined
\begin{flalign}
 M_{\rm m} = \frac{4\pi}{3}\bar{\rho}_{\rm m}(a)(1+\delta)R^3\,, \quad 
 M_{\rm bck,de} = \frac{4\pi}{3}\bar{\rho}_{\rm de}(a)R^3 \,. &&
\end{flalign}
In the previous expressions, $\bar{\rho}_{\rm m}(a)$ and $\bar{\rho}_{\rm de}(a)$ are the background matter and dark energy density, respectively, and $M_{\rm bck,de}$ represents the amount of dark energy included in a sphere with radius $R$. \cite{Pace2022} further extended the model to account also for dark energy perturbations.

It can be easily shown that Eq.~(\ref{eqn:eom_R}) leads to
\begin{flalign}\label{eqn:eom_delta}
 \delta^{\prime\prime} + &\, \left(2 + \frac{\eta}{H} +  \frac{H^{\prime}}{H}\right)\delta^{\prime} -  \frac{4}{3}\frac{{\delta^{\prime}}^2}{1+\delta} - \frac{3}{2}\Omega_{\rm m}(t)\delta(1+\delta) - && \nonumber\\
 & \left(\tilde{\sigma}^2-\tilde{\omega}^2\right)(1+\delta) - \frac{3\eta}{H}(1+\delta) = 0\,, &&
\end{flalign}
which contains the two additional terms $-\tfrac{3\eta}{H}(1+\delta)$ and $\frac{\eta}{H} \delta^{\prime}$. The latter is a proper friction term. At the linear level, the first term reads $-\tfrac{3\eta}{H}$ and leads to conceptual problems. It acts as a source term so that the particular solution is, for a model where $\eta=\eta_0H$ and for an EdS model, $\delta=-2\eta_0$. This implies that for $a\to 0$ an overdensity originates from an underdensity, or to put it in other terms, primordial perturbations are small but do not go to zero as expected.

As also noted in \cite{Carrilho2022}, this is due to the fact that dynamical friction depends only on the peculiar velocity of dark matter and that while for the overdensity we use comoving coordinates, for the evolution of the radius we use physical coordinates. In particular, the radius of the perturbation is the physical radius. When we go from the evolution of the density perturbation $\delta$ to that of $R$, the equation for the latter is \cite[see also][]{Carrilho2022}
\begin{flalign}\label{eqn:Rddot_correct}
 \frac{\mathrm{d}^2R}{\mathrm{d}t^2} = &\, -\frac{GM_{\rm m}}{R^2} - \frac{GM_{\rm bck,de}}{R^2}(1+3w_{\rm de}) &&\nonumber \\
 &\, - \frac{\sigma^2-\omega^2}{3}R - \eta\left(\frac{\mathrm{d}R}{\mathrm{d}t}-HR\right)\,. &&
\end{flalign}
This equation contains two terms originating from friction: one contribution dampens the collapse ($-\eta\,\mathrm{d}R/\mathrm{d}t$) and the other ($\eta HR$) enhances it. It is the term $\eta HR$ that exactly cancels the term $3\eta(1+\delta)/H$ in Eq.~(\ref{eqn:eom_delta}). From a physical point of view, the term in parenthesis, which can be considered analogous to a peculiar velocity, reconciles the two approaches, based on the overdensity and radius, respectively.

From now on, we will neglect the contribution of shear and rotation in our analysis and set them to zero, effectively considering an exact spherical symmetry, as we want to focus on the effects of friction and avoid an interplay among them and friction. The effects of shear and rotation on structure formation have been studied before for both smooth and clustering dark energy models and for different implementations, as already discussed in Sect.~\ref{sect:spcm}. We will not delve into them again, but we refer the reader to \cite{DelPopolo2013a,DelPopolo2013b,DelPopolo2013c,Pace2014b,Reischke2016a,Pace2017,Reischke2018,Pace2019b,Pace2022} for details.

In Sect.~\ref{sect:friction} we detail how to calculate the time dependence of $\eta$ with respect to time. Here, we simply summarise our findings. For the dynamical friction model of \cite{AntonuccioDelogu1994} we obtain $\eta = \eta_0\,H$ with $\eta_0$ constant, and for dissipative pressure friction we find $\eta = (\eta_1+\eta_2|P/P_0|^{\Gamma})\,H_0$, with $\eta_1$, $\eta_2$, and $\Gamma$ all constant.

\section{Theoretical models of friction}\label{sect:friction}
In this Section we describe the two models investigated in this work.

\subsection{Evolution of dynamical friction}\label{sect:friction1}

Here, we closely follow the works by \cite{AntonuccioDelogu1994} and \cite{DelPopolo2009}, to which we refer for more details.

In a scenario with hierarchical structure formation, structures of size $R$ form around the local maxima of the primordial density field smoothed over a scale of size $R$ \cite{Bardeen1986,Bond1988,Colafrancesco1989}\footnote{The peaks represent the fluctuations with positive density contrast in a Gaussian random field. Their properties can be calculated analytically following the theory of Gaussian random fields as outlined in the seminal work of \cite{Bardeen1986}. We note, however, that many works showed the development of non-Gaussianity during the early evolution of cosmic structures \cite{Fry1982,Fry1984,Fry1985,Fry1986,Bernardeau1992}. The Gaussian approximation is thus a very useful starting point to achieve analytical results which allow progressing in the subject and a clear understanding of the underlying physics.}. 
In this system, a test particle experiences a gravitational field which consists of a part associated with the smooth global mass distribution and a stochastic one which originates from the particle number fluctuations and gives rise to a frictional force $-\eta\mathbf{v}$ where $\eta$ is the dynamical friction coefficient and $\mathbf{v}$ the macroscopic velocity.

Using the virial theorem, it is rather straightforward to show that the dynamical friction coefficient reads
\begin{flalign}
 \eta = &\, \frac{4.44(Gm_{\rm a}n_{\rm a})^{1/2}}{N_{\rm tot}}\log{[1.12N_{\rm tot}^{2/3}]} && \nonumber \\
 = &\, 4.44\sqrt{\frac{3\Delta}{8\pi}}\frac{\log{[1.12N_{\rm tot}^{2/3}]}}{N_{\rm tot}}H = \eta_0\,H\,, &&
\end{flalign}
where we used the fact that $m_{\rm a}n_{\rm a}=\rho_{\rm m}=\bar{\rho}_{\rm m}\Delta$, with $\Delta$ the average overdensity of the perturbation. In principle, $\Delta$ would depend on the virialization recipe and on the cosmological model. However, here for simplicity, we will assume that it is constant, of the order of $\mathcal{O}(100)$.

This result shows that $\eta\propto H$ and that the value of $\eta_0$ depends on the total number of peaks $N$ within a protocluster, decreasing when $N$ increases. As discussed in \cite{AntonuccioDelogu1994}, $N$ is considered to be approximately constant for clusters. For $N_{\rm tot}=10^3$ we find $\eta_0 \approx 3\times 10^{-2}$, while for $N_{\rm tot}=10^4$ we have $\eta_0 \approx 4\times 10^{-3}$. Regardless of the exact value of $\eta_0$, the main conclusion is that $\eta_0\ll 1$.

Note also that this result is not in contrast with previous work, such as \cite{AntonuccioDelogu1994}, since an EdS model was assumed there. In fact, for such a model, neglecting radiation, we have $H\propto a^{-3/2}$. In fact, \cite{AntonuccioDelogu1992} showed that it is possible to associate a relaxation timescale with dynamical friction in galaxy clusters, and this is of the order of Hubble time, \textit{i.e.}, $1/\eta\propto H^{-1}$.

Finally, rather than expressing $\eta$ in terms of the total number of peaks $N$, it is worth using the mass of the perturbation $M$. Recalling that $M=m_{\rm a}N$, it is straightforward to obtain
\begin{flalign}\label{eqn:dyn_fric}
 \eta = 4.44\sqrt{\frac{3\Delta}{8\pi}}\frac{\log{[1.12(M/m_{\rm a})^{2/3}]}}{(M/m_{\rm a})}H = \eta_0\,H\,, &&
\end{flalign}
where we assume $m_{\rm a}=10^9\,h^{-1}\,M_{\odot}$. It is evident that $\eta_0$ is smaller for massive objects than for low-mass perturbations.

Eq.~\ref{eqn:dyn_fric} comes from Eqs. D3--D5 in \cite{DelPopolo2009}. Those equations come from the theory of stochastic forces in a gravitational field, as shown by \cite{Kandrup1980}. The terms $m_a$ and $n_a$ that determine the value of $\eta$ are related to the theory of Gaussian stochastic fields, valid for generic structures like galaxies, clusters, etc. In the text, we refer to the ``protoclusters'' just to estimate the values of these parameters. Again, in \cite{DelPopolo2009}, the equations and values quoted are used to study clusters of galaxies.

Following \cite{AntonuccioDelogu1994}, one can show that the peaks of the local density field with central height $\nu\ge\nu_{\rm c}$, with $\nu_{\rm c}$ a critical threshold, contribute to dynamical friction. The number of these objects can be calculated under the condition that the peak radius $r_{\rm pk}(\nu\ge\nu_{\rm c})$ is negligible with respect to the average peak separation $n_{\rm a}^{-1/3}(\nu\ge\nu_{\rm c})$.

The values required for the computation will necessarily depend on the particular cosmological model considered and on the filtering scale $R$. Here we do not pretend to derive exact values for the quantities above for every individual model we investigate in this work. Our aim is merely to estimate the order of magnitude of the relevant quantities, in particular the dynamical friction coefficient $\eta$ and to understand how this parameter affects the evolution of linear and nonlinear structures in $\Lambda$CDM and dark energy models.

Dynamical friction can be reinterpreted as a consequence of the fact that general relativity considers that a pressure term is not negligible within the dynamics dictated by the Friedmann equations. Hence, even if in Newtonian dynamics it is also possible to plug into the constraint equation for $H$ a pressure term, in fulfilment of energy conservation, the dynamical friction scenario ensures that friction is parametrised through the pressure itself, namely guaranteeing that the effects due to pressure are not subdominant than those inferred from the density alone. This generalises the previous scenario, where only density is involved, and proposes a way to better frame friction in time invoking that, from the general relativity recipe, the pressure effects are not negligible with respect to the density ones.

\subsection{Dissipative pressure friction}\label{sect:friction2}

Standard fluids can be grouped into two main classes: Newtonian and non-Newtonian. Clearly, the latter is found when the friction coefficient, $\eta$, depends on the variation of the shear velocity with respect to the radial coordinate. Our approach is simplified by virtue of the cosmological principle, \textit{i.e.}, the velocity field does not depend on radial coordinate. However, in this scenario, our friction coefficient is only a function of the Hubble rate. In other words, we do not emphasise the time dependence of it, assuming the simplest stationary case. It is clearly possible to generalise this case, including the effects due to the dynamics of the universe. Following the standard treatment, one can assume that the friction coefficient is a function of temperature and/or of thermodynamic quantities such as pressure, entropy, enthalpy and so forth. In the simplest case, we can assume what follows
\begin{flalign}\label{eqorl1}
    \eta=\left(\eta_1+\eta_2\left|\frac{P}{P_0}\right|^\Gamma\right)H_0\,, &&
\end{flalign}
where $\eta_1,\eta_2$ and $\Gamma$ are free real constants, with $P$ the total pressure of the universe and $P_0$ the total pressure today. The overall factor $H_0$ is just a normalisation factor that makes the two coefficients $\eta_1$ and $\eta_2$ dimensionless. From the second Friedmann equation,
\begin{flalign}\label{eqorl2}
    \frac{\ddot{a}}{a} = -\frac{4\pi G}{3}\left(\rho+\frac{3P}{c^2}\right)\,, &&
\end{flalign}
where $\rho$ and $P$ represent the total density and pressure and remembering that $\frac{\ddot{a}}{a} \equiv \dot{H}+H^2$, combining Eqs.~\eqref{eqorl1} and \eqref{eqorl2}, we infer
\begin{flalign}
 \eta = \left[\eta_1+\eta_2\left|\frac{2\dot{H} + 3H^2}{2\dot{H}_0+3H_0^2}\right|^\Gamma\right]H_0\,. &&
\end{flalign}

At this stage, we distinguish two cases, as reported in the following. 
\begin{itemize}
 \item[-] Slow-rolling approximation, namely $3H^2\gg 2\dot H$, leading to
  \begin{flalign}\label{eqapproxorl1}
    \eta = \left[\eta_1+\eta_2\left|\frac{H}{H_0}\right|^{2\Gamma}\right]H_0\,, &&
  \end{flalign}
reducing to our previous case, discussed above, with $\Gamma=\tfrac{1}{2}$, $\eta_1=0$ and $\eta_2=\eta_0$. This certifies that our previous approach turns out to be a suitable approximation that works well in a phase in which the effects of pressure are subdominant over those associated with the density only over the dynamics of the universe, although it does not account for the whole stages of its evolution.
\item[-] Fast-rolling approximation, namely $3H^2\ll 2\dot H$, having
 \begin{flalign}\label{eqapproxorl2}
   \eta = \left[\eta_1+\eta_2\left|\frac{\dot{H}}{\dot{H}_0}\right|^{\Gamma}\right]H_0\,, &&
 \end{flalign}
that shows a quite different case with respect to the one previously developed, where the dynamical effects, $\sim \ddot a$, are taken into account explicitly. 
\end{itemize}

In our computations, we investigate the consequences of Eq. \eqref{eqorl1}, prompting the main differences between the two limiting cases featured in Eqs. \eqref{eqapproxorl1} and \eqref{eqapproxorl2}. To this end, it appears essential to work out different sets of $\Gamma$ values that are not known \emph{a priori}. 

Following the standard treatment for gases, one can assume that the friction coefficient may become a function of pressure, resembling the temperature or entropy dependence of quantities as for standard gases. Even though this hypothesis appears valid for real gases, we here enable it for cold dark matter, under the hypothesis that its contribution remains small and does not modify significantly the approximation of collisionless dark matter. For the sake of completeness, in fact, dark matter is not fully collisionless, see e.g.
\cite{Bertone:2018krk}, and we may therefore expect a non-vanishing contribution provided by parameterizing the friction coefficient through the pressure as follows
\begin{flalign}
 \frac{\mathrm{d}\eta}{\mathrm{d}t} = \eta_2\,\Gamma\,H_0\, \left|\frac{P}{P_0}\right|^{\Gamma-2}\frac{P\dot{P}}{P_0^2}>0\,, &&
\end{flalign}
and then to determine the sign of $\dot P$, we can invoke Eq.~\eqref{eqorl2}, having $\dot{P} \propto -2(\ddot{H}+3H\dot{H})$ that can be rewritten as
\begin{flalign}
 \dot P=-2H^3(j-1)\frac{c^2}{8\pi G}\,, &&
\end{flalign}
recasting in terms of the jerk parameter, defined as $j\equiv -H^{-1}\dot q+2q^2+q$ \cite{Dunsby:2015ers}, with $q\equiv-1-\frac{\dot H}{H^2}$, the deceleration parameter and where we assumed that friction increases at large $z$.\footnote{Quite remarkably, as a thermodynamic fact, we ensure friction to increase with redshift in order to guarantee that at early times friction might be more relevant in analogy to standard thermodynamic systems, where collisions appear more relevant as pressure and temperature increase. Nevertheless, since the universe temperature tends to larger values, one presumes that friction increases as well, as a consequence of the stronger interactions among constituents. Clearly, the interaction itself appears subdominant with respect to the underlying fluids, mainly dominated by radiation for very large redshifts. Accordingly, even if radiation tends to dominate over the other fluids, the friction effects are not zero, evolving up to current time, where their contribution is no longer negligible, as matter and radiation magnitude dilute, becoming smaller than at early times, around $z\simeq 0$.}

The jerk parameter has been widely investigated in the literature \cite{Gruber:2013wua,Luongo:2011zz,Luongo:2013rba,Aviles:2016wel}, adding more information about the way the Hubble parameter would evolve w.r.t. cosmic time. In particular, in order to guarantee the Universe to pass through a matter- to dark energy- dominated phases, one requires the jerk parameter to be \emph{always} positive at late times. This may be seen evidently from the results obtained within the contexts of dark energy model-independent reconstructions, see e.g. \cite{Visser:2004bf,Carloni:2024zpl}, where this has been established. As a matter of remark, this occurrence allows us to conclude $\mathrm{d}H/\mathrm{d}t>0$ but not that dark energy today turns into a cosmological constant \cite{Luongo:2024fww}. Indeed, this may be argued instead if the jerk derivative would be well constrained, although this is currently not possible with current data.

Thus, since $\dot{H}>0$ and $j \geq 0$ \cite{Luongo:2015zgq,Capozziello:2022jbw}, we have  
\begin{flalign}
 \eta_2\Gamma<0\,, &&
\end{flalign}
that guarantees that $\dot \eta>0$, albeit it manifestly implies a degeneracy on the signs of these free constants. Indeed, since $\frac{dz}{dt}=-(1+z)H(z)$, we have
\begin{subequations}
 \begin{flalign}
  \eta^{\prime} \simeq &\, 2\tilde{\eta}_0(1+z)^3P_{\rm rad}^{\Gamma-1}\,, &&\\
  \eta^{\prime} \simeq &\, 0\,, &&
 \end{flalign}
\end{subequations}
respectively, for an Einstein-de Sitter universe dominated by radiation and matter, where $\tilde{\eta}_0\equiv 2H_0\eta_2\Gamma$ and $\eta^{\prime}\equiv \frac{\mathrm{d}\eta}{\mathrm{d}z}$. Consequently, in view of the above results, we can take $\eta_2 < 0$ and $\Gamma>0$ or vice versa, having moreover 
\begin{subequations}
 \begin{flalign}
  \eta_1 = &\, 0\,,\, \textrm{and}, \, \eta_2 \in[0,1]\,, \Gamma\in[-2;-1]\,, \qquad {\rm or} &&\\
  \eta_1 = &\, 0\,,\, \textrm{and}, \, \eta_2 \in[-1,0]\,, \Gamma\in[1;2]\,, &&
\end{flalign}
\end{subequations}
where $\eta_1$ is arbitrarily imposed to vanish as it appears unessential in determining $\eta$ and the values $\pm 1$ selected for normalizing $\eta_2$ and $\Gamma$. These priors are not directly arguable from observations, but rather they match our current expectations over the cosmographic series inferred at late times, see e.g. \cite{Cattoen:2008th,Visser:2009zs}.

Having the above, case $\Gamma=0$ corresponds to a constant friction term, while case $\eta_2=0$ implies a frictionless model. Consequently, we argue that 
\begin{itemize}
 \item[-] the first case described above implies a model where friction acts attractively, \textit{i.e.}, the contribution acts as a standard fluid within the Zeldovich limit. Hence the total friction, taking also into account the cosmological expansion is increased,
 \item[-] the second case described above implies a model where friction acts repulsively, \textit{i.e.}, the sign is compatible with the dark energy equation of state. We will not consider this case here, as it does not represent the ordinary behavior of friction on matter.
\end{itemize}
From the aforementioned considerations, we fit our models within the intervals of priors determined above. We discuss our findings in what follows.

\section{Evolution of linear perturbations}\label{sect:LinPert}

In this Section we discuss the evolution of linear perturbations and show how they are affected by dynamical friction.

Although small, $\eta$ affects the initial conditions of Eq.~(\ref{eqn:delta_L}), similarly to what happens in early dark energy and massive neutrinos scenarios. This happens because the friction term $\eta/H$ is small but not negligible. To elucidate this concept, let us denote by $\eta_{\rm ini} = \eta(a_{\rm ini})/H(a_{\rm ini})\ll 1$ the value to be used at an initial time to evaluate the initial conditions. In general, at an early time, a given cosmological model is well approximated by an EdS model, so that we can assume $\Omega_{\rm m}(t)\approx 1$ and $H^{\prime}/H\approx -3/2$ and that the growth factor follows a power-law behavior, $\delta=Aa^n$, where $A$ is an arbitrary constant amplitude that can be safely set to one. Plugging the Ansatz for $\delta$ into Eq.~(\ref{eqn:delta_L}), we need to solve a second order algebraic equation for the exponent $n$:
\begin{flalign}
 n^2 + \left(\frac{1}{2} + \eta_{\rm ini}\right)n - \frac{3}{2} = 0\,, &&
\end{flalign}
whose growing mode solution is, for $\eta_{\rm ini}\ll1$, $n_{+} = 1 - 2\eta_{\rm ini}/5$ if curvature and radiation are neglected. This result is expected from physical considerations: friction slows down the collapse and $n_{+}<1$ when $\eta\neq 0$. We can compare this result with a generic early dark energy model. If we denote by $\Omega_{\rm e}\ll 1$ the amount of early dark energy, we have $\Omega_{\rm m}= 1-\Omega_{\rm e}$ and find for the growing mode $n_{+} = 1 - 3\Omega_{\rm e}/5$. This same expression applies to massive neutrinos, if we replace $\Omega_{\rm e}$ with the fraction of massive neutrinos $f_{\nu} = \Omega_{\nu}/\Omega_{\rm m}$, with $\Omega_{\nu}$ the neutrinos density parameter. Note the similarity of the different physical systems. In this case, too, therefore, structure formation is slowed down. This means that to match the growth rate of structures in the $\Lambda$CDM model, the initial amplitude of the perturbations must be higher. We will discuss this in more detail when investigating the evolution of nonlinear perturbations.

After this discussion, we can now study the evolution of the growth factor. We consider two clustering species, baryons (denoted with the subscript $\mathrm{b}$) and dark matter (denoted with the subscript $\mathrm{cdm}$), of which only the latter is affected by dynamical friction. The two equations to be solved are
\begin{subequations}
 \begin{flalign}
  \delta_{\rm b}^{\prime\prime} + \left(2 + \frac{H^{\prime}}{H}\right)\delta_{\rm b}^{\prime} - \frac{3}{2}\left(\Omega_{\rm b}\delta_{\rm b} + \Omega_{\rm cdm}\delta_{\rm cdm}\right) = &\, 0\,, &&\\
  \delta_{\rm cdm}^{\prime\prime} + \left(2 + \frac{\eta}{H} + \frac{H^{\prime}}{H}\right)\delta_{\rm cdm}^{\prime} - \frac{3}{2}\left(\Omega_{\rm b}\delta_{\rm b} + \Omega_{\rm cdm}\delta_{\rm cdm}\right) = &\, 0\,, &&
 \end{flalign}
\end{subequations}
and we define the total growth factor as
\begin{flalign}\label{eqn:tgf}
 \delta_{\rm m} = \frac{\Omega_{\rm b}\delta_{\rm b} + \Omega_{\rm cdm}\delta_{\rm cdm}}{\Omega_{\rm b}+\Omega_{\rm cdm}}\,. &&
\end{flalign}

\subsection{Initial settings}

If we consider a single clustering species, we can assume that at early times $\delta = A\,a^n$, where both $A$ and $n$ are constant. This leads to a second-order algebraic equation for $n$ whose solution, for $\eta=\eta_0H$ and $\eta_0\ll 1$, is $n=1-\tfrac{2}{5}\eta_0$. We can proceed in a similar way if we consider both dark matter and baryons. In this case, we can also assume a power law such that $\delta_{\rm b}=A\,a^n$ and $\delta_{\rm cdm}=B\,a^n$, with constant $A$, $B$, and $n$ to determine. We assume that the two perturbations have the same slope but different amplitudes. If we plug this Ansatz in the previous equations, we obtain, taking into account that at early times the background is extremely well represented by an EdS model,
\begin{subequations}
 \begin{flalign}
  n^2\delta_{\rm b} + \frac{1}{2}n\delta_{\rm b} - \frac{3}{2}(\Omega_{\rm b}\delta_{\rm b}+\Omega_{\rm cdm}\delta_{\rm cdm}) = &\, 0\,, &&\\
  n^2\delta_{\rm cdm} + \left(\frac{1}{2}+\eta_{\rm ini}\right)n\delta_{\rm cdm} - \frac{3}{2}(\Omega_{\rm b}\delta_{\rm b}+\Omega_{\rm cdm}\delta_{\rm cdm}) = &\, 0\,.&&
 \end{flalign}
\end{subequations}

From the first equation we can solve for the Poisson term and plug it into the second equation, in order to find a relation between the initial perturbations
\begin{flalign}\label{eqn:ratio}
 \frac{\delta_{\rm b}}{\delta_{\rm cdm}} = 1 + \frac{\eta_{\rm ini}}{n+1/2}\,, &&
\end{flalign}
which shows that the baryon perturbation amplitude is increased with respect to that of cold dark matter, as baryons are not affected by friction. Also note that for vanishing friction, $\delta_{\rm b}=\delta_{\rm cdm}=\delta_{\rm m}$, as expected.

Plugging Eq.~(\ref{eqn:ratio}) into the equation for $\delta_{\rm cdm}$, leads to an equation for the exponent $n$:
\begin{flalign}
 n^2 + \left(\frac{1}{2}+\eta_{\rm ini}\right)n - \frac{3}{2}\left(1+\Omega_{\rm b}\frac{\eta_{\rm ini}}{n+1/2}\right) = 0\,. &&
\end{flalign}
Note that when $a\ll 1$, $\Omega_{\rm b}=\Omega_{\rm b,0}/\Omega_{\rm m,0}$, with $\Omega_{\rm m,0}=\Omega_{\rm b,0}+\Omega_{\rm cdm,0}$. Hence, the term which involves baryons is always constant and subdominant with respect to unity. For realistic values of both $\Omega_{\rm b,0}$ and $\Omega_{\rm m,0}$, we have $\Omega_{\rm b}$ of the order of 0.2.

Although it is possible to rewrite it as a third-order equation, it is simpler to solve it as is, to avoid going through complex radicals. However, we can use the cubic equation to investigate the behavior of the initial slope when $\eta_{\rm ini}\ll1$. The equation to be solved is
\begin{flalign}
 n^3 + (1+\eta_{\rm ini})n^2 - \frac{1}{4}(5-2\eta_{\rm ini})n - \frac{3}{4}\left(1+2\Omega_{\rm b}\eta_{\rm ini}\right) = 0\,. &&
\end{flalign}
If $\eta_{\rm ini}=0$, it is easy to see that the solutions are $n=-\tfrac{1}{2}$, $n=-\tfrac{3}{2}$ and $n=1$, of which only the latter two are the physical ones, corresponding to the decaying and growing modes, respectively. At first order in $\eta_{\rm ini}$, we obtain $n=1-\tfrac{2}{5}\eta_{\rm ini}(1-\Omega_{\rm b})$, where it is understood that $\Omega_{\rm b}$ is evaluated at early times. The term in parentheses is positive for realistic values of the baryon density, and hence, also in this case, we can see that friction dampens structure formation. If we set, for simplicity $1-\Omega_{\rm b}\approx 1$, we recover the previous result, showing the excellent agreement between the two approaches.

Since the two equations of motion are different, we cannot combine them into a single equation for the variable $\delta_{\rm m}$. Therefore, to highlight the importance of friction and the differences induced by having two species rather than one, we consider the evolution of the linear growth factor $D_{+}$ as a function of the redshift $z$ considering either one or two species. As differences among the two different cases are of the order of $0.1\%$, for simplicity, we will consider just the case where both dark matter and baryons are affected by dynamical friction. This will also simplify the discussion when considering clustering dark energy models.

To justify the Ansatz previously used for the time evolution of baryons and cold dark matter, let us consider the two equations describing the evolution of perturbations in the standard case, when no friction contribution is present. In this case, the two equations are the same, and considering the definition of $\delta_{\rm m}$, the combined equation has the same form as the previous two, but with the term $\Omega_{\rm b}\delta_{\rm b} + \Omega_{\rm cdm}\delta_{\rm cdm}$ replaced by $\Omega_{\rm m}\delta_{\rm m}$. Hence, it is clear that the two variables evolve in the same way but have different amplitudes. As the friction term is very small, we can assume that this Ansatz is still valid in our case. This is certainly true for the equations that govern the growth factor, but similar considerations can be made later when discussing the equation for the spherical collapse model.

The qualitative discussion above can be made more quantitative following \cite{Bird2020}. If we consider the difference $\delta_{\rm cb} = \delta_{\rm cdm}-\delta_{\rm b}$, this quantity decays as $a^{-1/2}$ in the EdS regime, showing that the two quantities will evolve in the same way and we can thus assume that they only differ in their amplitude when setting the initial conditions. For a more detailed treatment, we refer the reader to \cite{Bernardeau2002,Schmidt2016}.

\subsection{Growth factor}

In this Section we present the effects of friction on the evolution of the linear growth factor, using the two recipes described above. We first consider its effect on the $\Lambda$CDM reference model. We then compare it with six different dark energy models, two of them characterized by a constant equation of state, $w_{\rm de} = -0.9$ and $w_{\rm de} = -1.1$ (DE1 and DE2), and four with a linearly time-dependent equation of state, $w_{\rm de} = w_0 + w_a(1-a)$, with $w_0$ and $w_a$ constant \citep{Chevallier2001,Linder2003}. We choose the following combinations of parameters $(w_0, w_a)$: $(-0.9, \pm 0.3)$ (DE3 and DE4) and $(-1.1, \pm 0.3)$ (DE5 and DE6) to cover quintessence, phantom and barrier crossing models.\footnote{These are the same parameters adopted for the DEMNUni simulations \citep{Carbone2016}, in which we tested the implementation of our spherical collapse code in case of absence of friction contributions to show a very good quantitative agreement between the theoretical and numerical halo mass function.}

\begin{figure*}[!ht]
 \centering
 \includegraphics[scale=0.35]{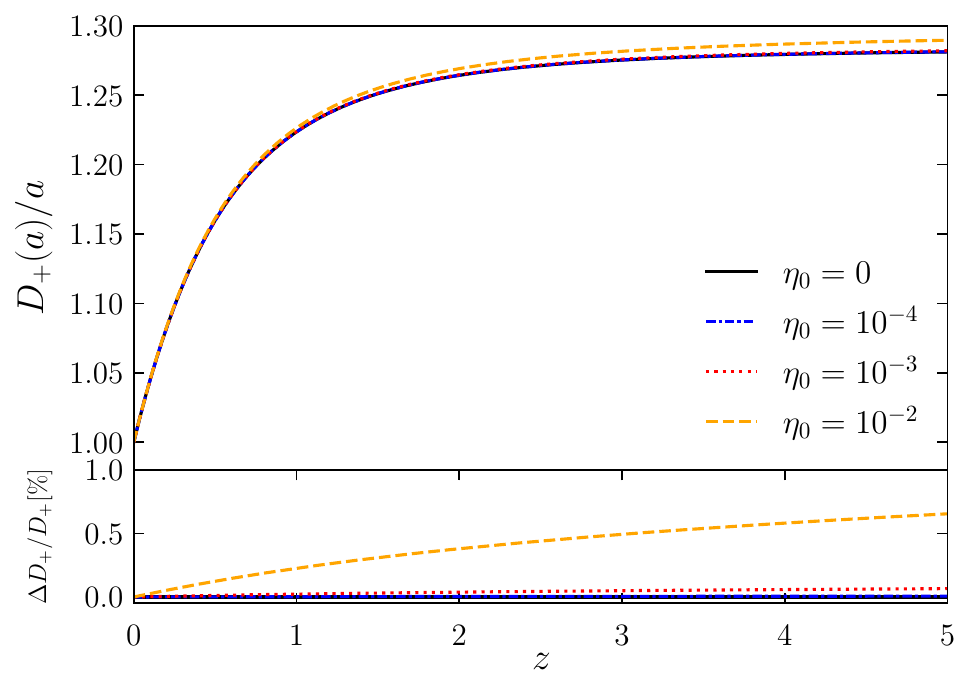}
 \includegraphics[scale=0.35]{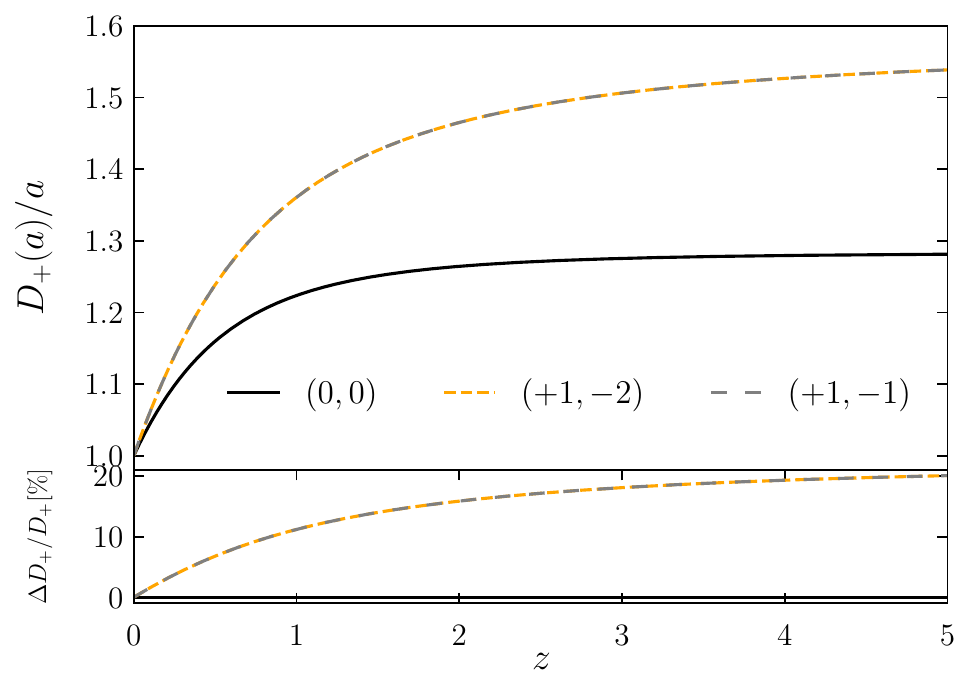}
 \caption{Evolution of the growth factor for the $\Lambda$CDM model for the two recipes used in this work to describe cosmic friction. Left panel: dynamical friction as parametrised in Eq.~(\ref{eqn:dyn_fric}). Right panel: dissipative pressure friction as parametrised in Eq.~(\ref{eqorl1}). Different colours and line styles refer to different values of the free parameters.}
 \label{fig:gf_LCDM}
\end{figure*}

In Fig.~\ref{fig:gf_LCDM} we present the evolution of the linear growth factor, divided by the scale factor, for the $\Lambda$CDM model, showing results for both dynamical friction (left panel) and dissipative pressure friction (right panel). The considerations which we are going to make for this model will also apply to the other models studied in this work.

As $\eta$ opposes the formation of structures, perturbations grow more slowly with respect to a model where the friction term is absent. Hence, because we want the same rate today (all the curves have been normalised to unity today), structures need to grow faster in the past. This explains the differences observed. Note that this behavior is analogous to $\Lambda$CDM versus EdS. Therefore, the higher the value of $\eta$, the higher the difference from a model without friction.

For dynamical friction, in general, differences are very small and only appreciable for the highest value chosen $\eta_0 = 10^{-2}$. The effects of dissipative pressure friction follow the same trend. Dissipative pressure friction behaves similarly to dynamical friction, and when the models are all normalized to one today, they present an increase in the growth factor. These considerations explain the behavior found in Fig.~\ref{fig:gf_LCDM}. We also note that changing $\Gamma$ has no effect on the evolution of perturbations, but this is only true for the $\Lambda$CDM model, where density and hence pressure do not evolve in time. Finally, the effects induced by dissipative pressure friction are much larger than those induced by dynamical friction. This happens because whilst the contribution of dynamical friction is constant in time, dissipative friction changes with time and its effects are stronger later in the cosmic evolution.

\begin{figure*}
 \centering
 \includegraphics[scale=0.35]{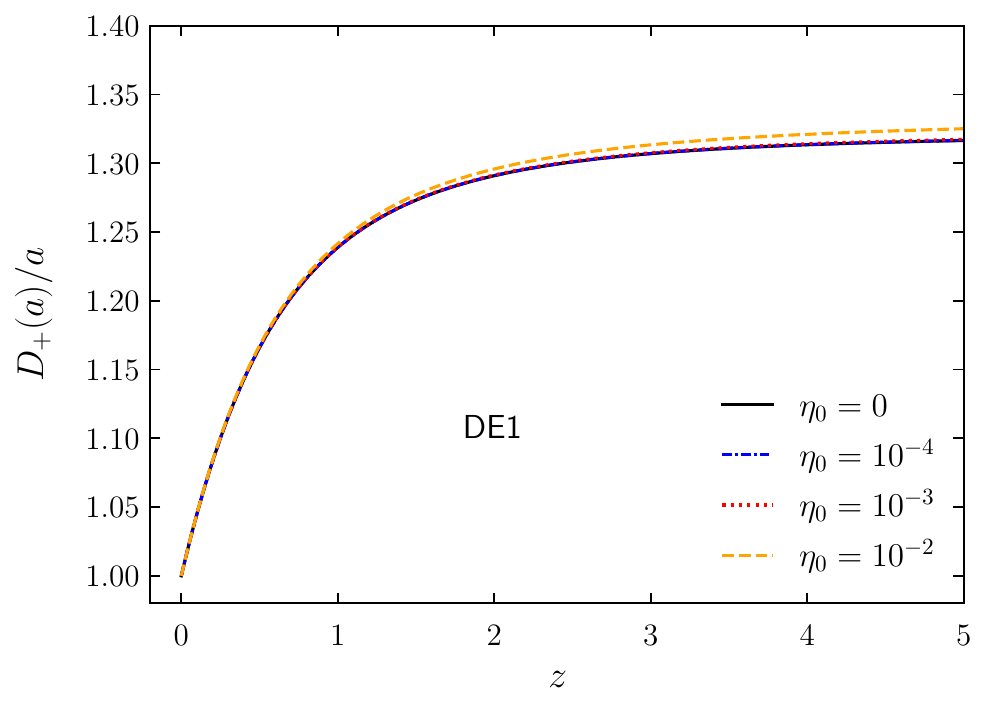}
 \includegraphics[scale=0.35]{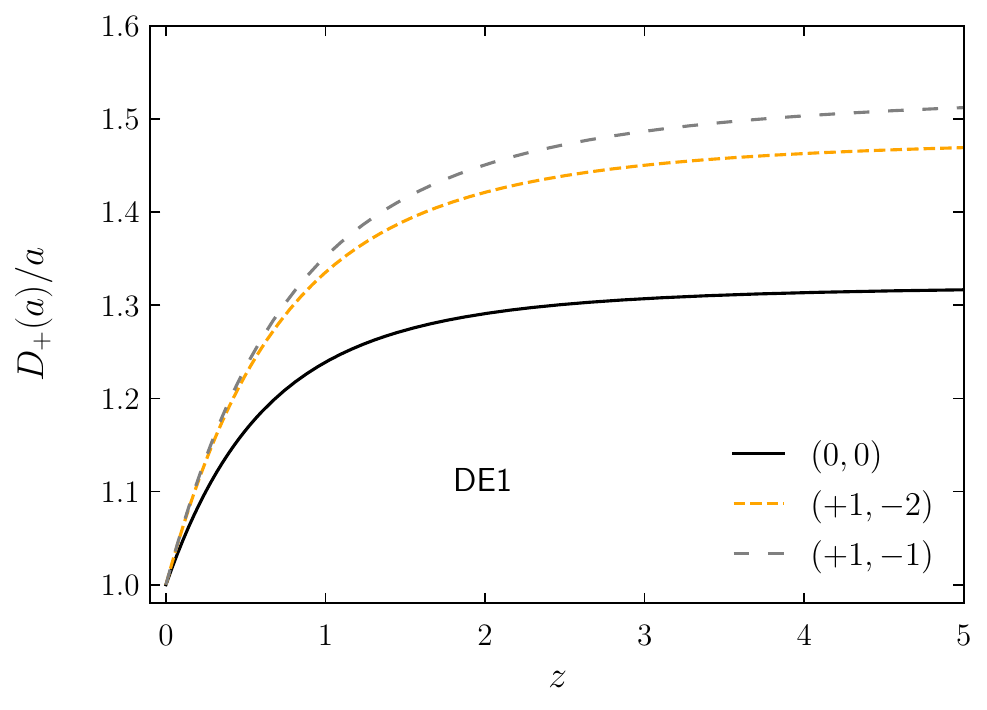}
 \caption{Evolution of the growth factor for the model DE1 for dynamical friction (left panel) and dissipative pressure friction (right panel). Line styles and colours are as in Fig.~\ref{fig:gf_LCDM}.}
 \label{fig:gf_wCDM}
\end{figure*}

Similar considerations can be made for Fig.~\ref{fig:gf_wCDM} where we show the effects of dynamical and dissipative friction for the model DE1. We do not show the other models because the qualitative behavior is the same and they only differ quantitatively. This applies to all the quantities we will present for dark energy models. The effect of dynamical friction for these models is similar to the $\Lambda$CDM model when we consider the relative difference with respect to a model without friction. This is clearly understood, as dark energy only affects the background and the dynamical equation to be solved is the same for both $\Lambda$CDM and the dynamical dark energy models. Furthermore, the correction due to $\eta_0$ is constant in time.

A deeper discussion on the effect of dissipative pressure friction is worth pursuing. In this case, because pressure changes over time, the parameter $\Gamma$ has an effect and the higher it is (in absolute value), the higher the difference with respect to the model not affected by friction. For the $\Lambda$CDM model, a positive $\eta_2$ implies a higher growth factor. In addition, because pressure changes over time, the observed effect depends on the particular equation of state. Models DE1, DE3, DE4 and DE5 share similar effects for the cases $(+1,-2)$ and $(+1,-1)$ while for model DE2 differences start to increase. Model DE6 presents the strongest effect for the combination $(+1,-2)$ and this is due to the fact that the equation of state remains in the phantom regime all the time and, as a consequence, the pressure is higher (in absolute value).

\section{Evolution of nonlinear perturbations}\label{sect:NonLinPert}
The study of nonlinear perturbations is based on the approach of the spherical collapse, whose equations have been derived in Sect.~\ref{sect:spcm}. We consider once again two perturbed fluids, dark matter and baryons, where only the first experiences the effects of dynamical friction. To derive the initial conditions of the full nonlinear equation we consider that the averaged matter density $\delta_{\rm m}$ diverges \cite{Pace2010}. However, for numerical stability, we do not use $\delta_{\rm m}$, but rather its inverse, so the collapse takes place when $1/\delta_{\rm m}\to 0$, as discussed in \cite{Pace2017a}.

Some important quantities which can be defined within the formalism of spherical collapse are the turn-around scale factor, $a_{\rm ta}$ and the nonlinear overdensity at turn-around $\zeta$. The first represents the time at which the spherical perturbation reaches its maximum extension before collapsing and shrinking, while the latter is defined as $\zeta = 1+ \delta_{\rm NL}(a_{\rm ta})$. The nonlinear overdensity at turn-around is constant for an Einstein-de Sitter model, but for any other model, including the $\Lambda$CDM one, changes in time. We verified that dynamical friction has either a negligible or very small effect on both the turn-around scale factor $a_{\rm ta}$ and the nonlinear overdensity at turn-around, $\zeta$. When considering dissipative pressure friction, instead, the value of $a_{\rm ta}$ changes up to $2\%$, while $\zeta$ differs up to $15\%$ with respect to a model without friction.

\begin{figure*}
 \centering
 \includegraphics[scale=0.35]{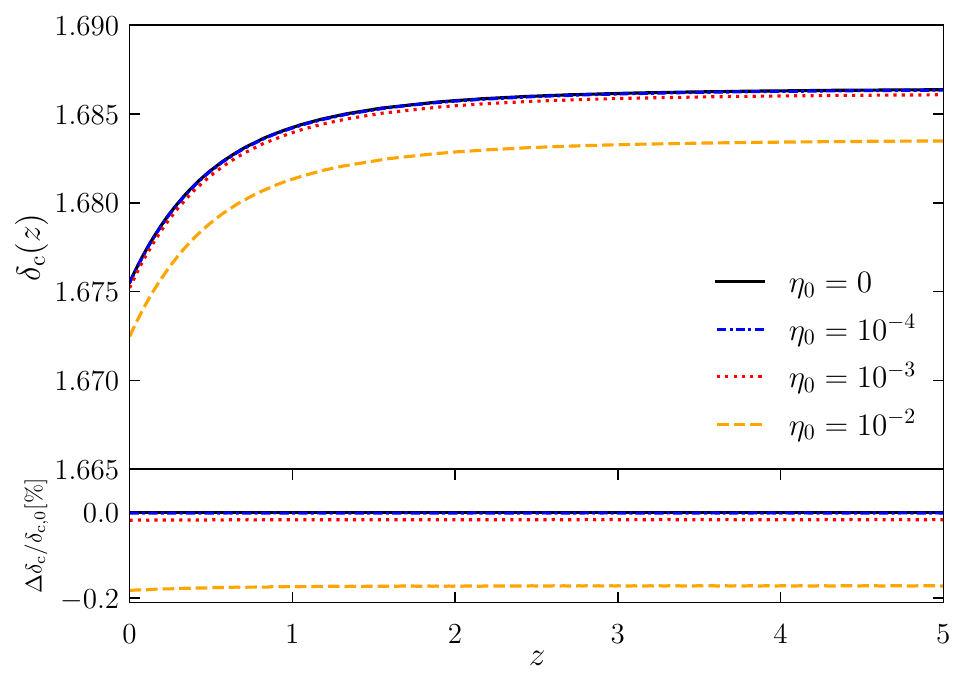}
 \includegraphics[scale=0.35]{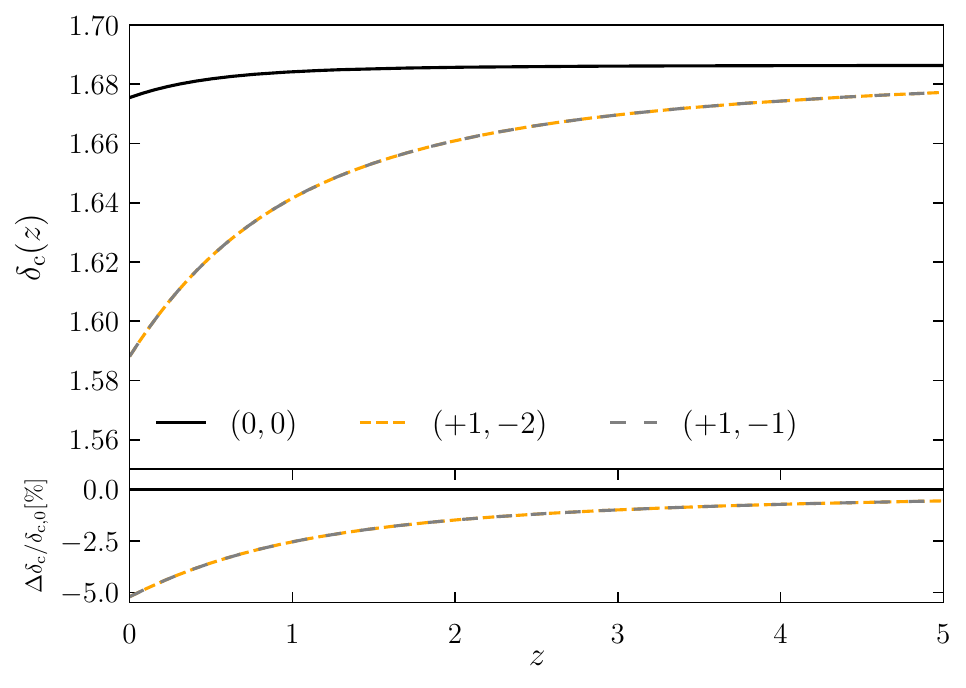}
 \caption{Evolution of the linearly extrapolated overdensity $\delta_{\rm c}$ for the $\Lambda$CDM model. In the left (right) panel we show results for the dynamical (dissipative pressure) friction. Line styles and colours are as in Fig.~\ref{fig:gf_LCDM}.}
 \label{fig:deltaC_LCDM}
\end{figure*}

Two additional and very important quantities, also from an observational point of view, are the linearly extrapolated overdensity $\delta_{\rm c}$ and the virial overdensity $\Delta_{\rm vir}$. Linear overdensity $\delta_{\rm c}$ is, together with the linear growth factor, the main ingredient of the halo mass function which, as the name says, represents the linear evolution of matter density perturbations, whose initial conditions are found to have the nonlinear matter density evolution to diverge at collapse time. The linear overdensity thus represents the solution of a linear differential equation but with initial conditions derived from its nonlinear counterpart. Its time evolution as a function of the collapse redshift $z$ for the $\Lambda$CDM model is presented in Fig.~\ref{fig:deltaC_LCDM}. We show the results for the dark energy models in \ref{sect:DE} in Fig.~\ref{fig:deltaC_wCDM_DE1}, to which we refer for a discussion of our findings.

Although the overdensity $\delta_{\rm c}$ is a linear quantity, the differences between the models are larger than with respect to the growth factor, despite the fact that we solve the same differential equation. This happens because for the growth factor the initial conditions are the same for all the models, while for $\delta_{\rm c}$ they are model dependent and are obtained by first solving a nonlinear differential equation. Focusing again on the $\Lambda$CDM model, for the dynamical friction case, we see that, as expected, differences are enhanced when the free parameter $\eta_0$ increases, so that for $\eta_0 = 0.01$ the linear overdensity is about 0.2\% lower than when $\eta_0=0$, regardless of the collapse redshift considered. The highest difference is at $z=0$ and at a higher redshift all models recover the corresponding value for the EdS model. When friction becomes stronger, $\delta_{\rm c}$ is lower, even though the initial overdensity is larger. This is because to collapse at the same time as a model without friction, perturbations require a lower threshold for the collapse to compensate the presence of friction.

In the right panel of Fig.~\ref{fig:deltaC_LCDM} we consider dissipative pressure friction applied to the $\Lambda$CDM model. As expected, the effects induced by friction are higher than for dynamical friction and, analogously to the growth factor, the parameter $\Gamma$ is irrelevant (but it will not be for dynamical dark energy models). The differences, now, are of the order of $-5\%$ and $\delta_{\rm c}$ decreases, in agreement with what was found above. However, this time, for all combinations of parameters, we recover the standard $\Lambda$CDM model at higher redshift. This is expected as for dissipative pressure friction the pressure term grows with time and is, therefore, important at relatively late times, when the dark energy component dominates.

The virial overdensity $\Delta_{\rm vir}$ represents the value that the nonlinear overdensity reaches at virialization. To determine how virialization proceeds, we follow the same procedure as in \cite{Carrilho2022} for interacting dark energy models. Even though this is not the physical case we are studying here, for smooth dark energy the equations of motion are the same, and dark matter experiences, in both cases, an additional dynamical friction force.

If we denote with $K$ the total kinetic energy of the system and with $W$ its potential energy, the expression for the virial theorem is modified as
\begin{flalign}\label{eqn:VT}
 2\langle K \rangle + \langle W \rangle + \sum_{i}\langle \mathbf{F}_{i}^{\rm fric} \cdot \mathbf{r}_{i} \rangle = 0\,, &&
\end{flalign}
where $\mathbf{F}_{i}^{\rm fric}$ is a non-conservative force, $\mathbf{r}_{i}$ the physical position of particles over which the summation is extended to. The friction force can be written as
\begin{flalign}
 \mathbf{F}_{i}^{\rm fric} = -m\eta\, x_i^{\prime}\,, &&
\end{flalign}
with $m$ the mass of individual particles.

We can make explicit the different terms entering into the virial theorem. To do so, we follow the notation of \cite{Cataneo2019,Carrilho2022} and denote with\footnote{We corrected a typo in the relation between $\delta$ and $y$ in Eq.~(28) of \cite{Carrilho2022} and Eq.~(34) of \cite{Cataneo2019}.}
\begin{flalign}
 E_0 = \frac{3}{10}M_{\rm m}(H_0\,R_i)^2\,, \qquad 
 \delta = (1+\delta_i)\,y^{-3} - 1\,, &&
\end{flalign}
so that
\begin{subequations}
 \begin{flalign}
  K = &\, \frac{H^2}{H_0^2}\left[\frac{a}{a_i}(\partial_Ny+y)\right]^2\,E_0\,, && \\
  W_{\rm N} = &\, -\Omega_{\rm m,0}\frac{1}{a\,a_i^2}\,y^2(1+\delta)\,E_0\,, && \\
  W_{\rm de} = &\, -\frac{H^2}{H_0^2}(1+3w_{\rm de})\Omega_{\rm de}(a)\left(\frac{a}{a_i}\right)^2y^2\,E_0\,, && \\
  W_{\rm fr} = &\, -2\frac{\eta}{H}\frac{H^2}{H_0^2}\left(\frac{a}{a_i}\right)^2y\,\partial_Ny\,E_0\,, &&
 \end{flalign}
\end{subequations}
with $N \equiv \ln{a}$, and where we decomposed the total potential energy $W$ in its contributions, \textit{i.e.}, $W = W_{\rm N} + W_{\rm de} + W_{\rm fr}$, with $W_{\rm N}$ the Newtonian potential due to matter perturbations, $W_{\rm de}$ the background dark energy contribution and $W_{\rm fr}$ the friction contribution. In a more familiar form, expressing these quantities in term of the physical radius $R$, we have
\begin{subequations}
 \begin{flalign}
  K = &\, \frac{3}{10}M_{\rm m}\dot{R}^2\,, && \\
  W_{\rm N} = &\, -\frac{3}{5}\frac{GM_{\rm m}^2}{R}\,, &&\\
  W_{\rm de} = &\, -\frac{3}{5}\frac{GM_{\rm m}M_{\rm bck,de}}{R}(1+3w_{\rm de})\,, &&\\
  W_{\rm fr} = &\, -\frac{3}{5}\eta\,M_{\rm m}R\left(\dot{R}-HR\right)\,. &&
 \end{flalign}
\end{subequations}
Note how in the definition of $W_{\rm fr}$, where we use the physical radius, there appears once again the term $\dot{R}-HR$ which was discussed above.

\begin{figure*}
 \centering
 \includegraphics[scale=0.35]{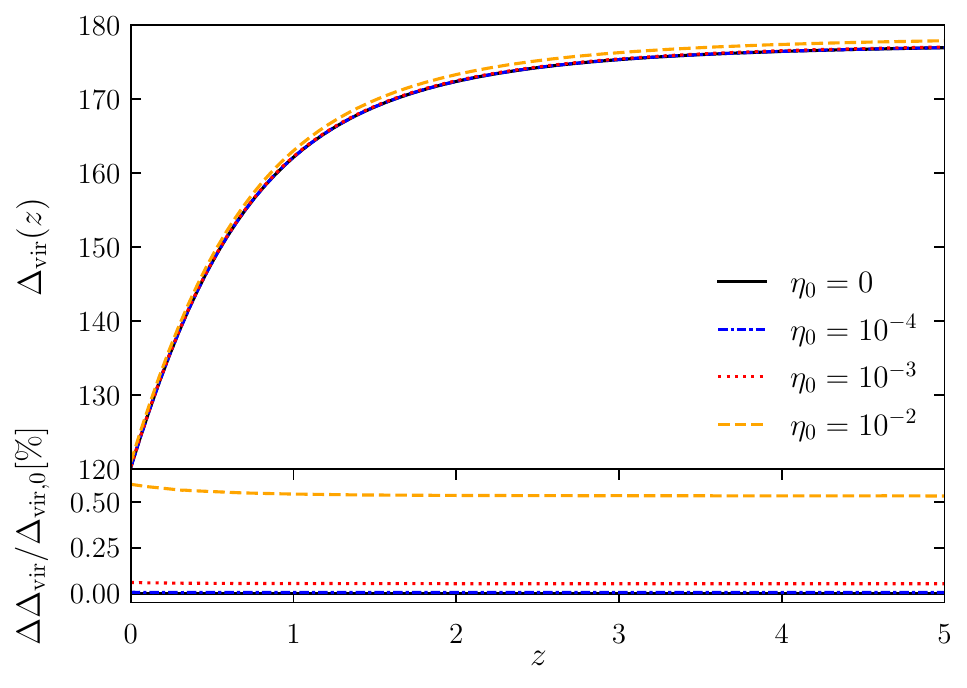}
 \includegraphics[scale=0.35]{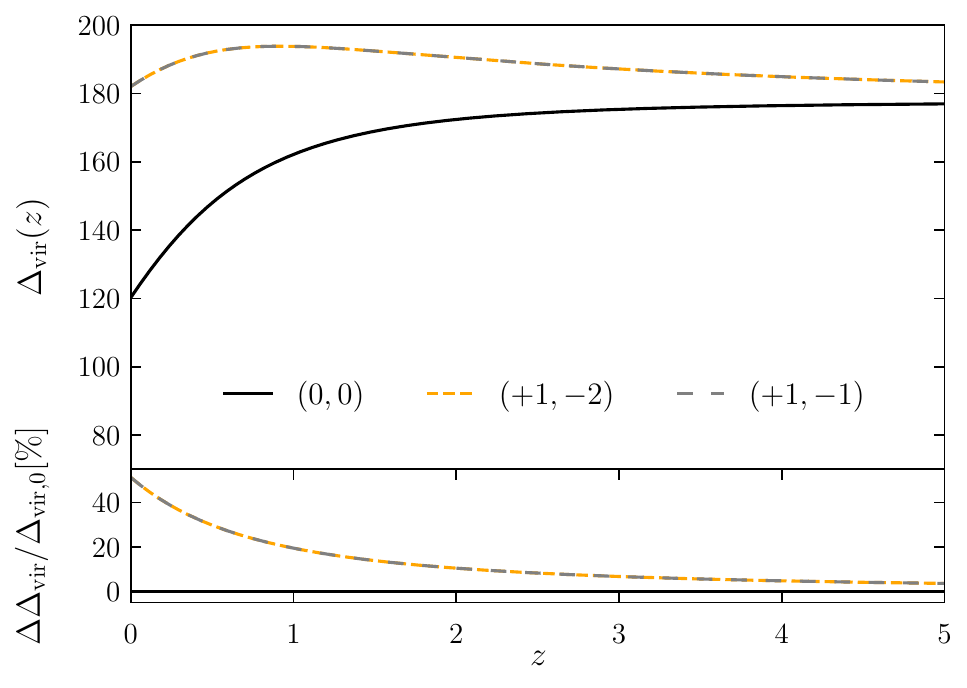}
 \caption{Evolution of the virial overdensity $\Delta_{\rm vir}$ for the $\Lambda$CDM model. In the left (right) panel we show results for the dynamical (dissipative pressure) friction. Line styles and colours are as in Fig.~\ref{fig:gf_LCDM}.}
 \label{fig:DeltaV_LCDM}
\end{figure*}

Therefore, according to Eq.~(\ref{eqn:VT}), to find the virialization epoch, we solve the equation
\begin{flalign}
 2K + W_{\rm N} + W_{\rm de} + W_{\rm fr} = 0\,, &&
\end{flalign}
and the virial overdensity is defined as
\begin{flalign}\label{eqn:DeltaVir}
 \Delta_{\rm vir} = (1+\delta_{\rm NL})\left(\frac{a_{\rm c}}{a_{\rm vir}}\right)^3\Omega_{\rm m}(a_{\rm vir})\,, &&
\end{flalign}
with $a_{\rm c}$ and $a_{\rm vir}$ the collapse and virialization scale factor, respectively. The nonlinear overdensity $\delta_{\rm NL}$ is evaluated in the virialization epoch.

Knowing the virialization epoch, using the formalism outlined above, we can also determine the ratio of the virial radius to the turn-around radius, $R_{\rm vir}/R_{\rm ta}$. This quantity gives an idea of how much the perturbation shrinks from its maximum size once it reaches the virial equilibrium. This quantity is, therefore, affected by friction, but it is also cosmology- and time-dependent. It is constant in time only for an EdS model, for which $R_{\rm vir}/R_{\rm ta} = 1/2$ (in the absence of friction contributions). Once again, the effect due to dynamical friction is extremely small, but it is more relevant, about a few percent for dissipative pressure friction.

We can now discuss the effects of friction on the evolution of virial overdensity $\Delta_{\rm vir}$, as defined in Eq.~(\ref{eqn:DeltaVir}) and presented in Fig.~\ref{fig:DeltaV_LCDM} for the $\Lambda$CDM model. In the left (right) panel we consider dynamical (dissipative pressure) friction. Once more, the first case leads to very small effects, except for $\eta_0 = 10^{-2}$, with differences of about $0.5\%$, relatively constant in time, with a small increase at late times when the cosmological constant dominates. For the case of dissipative pressure friction, the effects are relevant, as extensively discussed before, reaching up to $50\%$ today. These results are in perfect agreement with our previous analysis of the different nonlinear quantities, in particular the overdensity at the turn-around $\zeta$.

In conclusion, dynamical friction appears to have only small signatures in the models we investigated. On the other side, albeit theoretically motivated, the dynamical pressure model shows highly-performing changes into the shapes that are expected once friction is involved. Thus, on the one hand, the latter clearly appears more probable to be measured. Conversely, the first case appears to be hard to disentangle from the background dark energy. Remarkably, both are not excluded since it would be possible to experimentally constrain the associated free parameters to check which values are thus expected to correctly reproduce a friction contribution. For the sake of completeness, we expect the dynamical pressure model to behave less deeply, being constructed in the simplest way in which the pressure itself is modified as a result of the inclusion of a friction term.

\section{Effects of the modified evolution of the mass function}\label{sect:MassFunction}

In this section, we build upon results of previous sections, and we investigate the halo mass function. This is an important observable that is used to probe cosmology at late times by counting isolated bounded halos in the Universe. As in the case of the CMB or the matter power spectrum, the halo mass function depends on the cosmological model assumed.

From a theoretical point of view, the functional form of the halo mass function can be inferred from fits to $N$-body simulations \citep{Jenkins2001,Tinker2008,Watson2013}, although analytical derivations exist \citep{Press1974,Sheth2001}. In this work, we make a very conservative choice and use the Sheth-Tormen (ST) halo mass function \citep{Sheth2001}. This choice is dictated by the fact that \cite{Creminelli2010} used this formulation to describe the halo mass function in clustering dark energy models with null sound speed. We will come back to this point later.

The halo mass function is
\begin{flalign}
 \frac{\mathrm{d}n}{\mathrm{d}M} = \frac{\bar{\rho}_{\rm,0}}{M^2}f(\nu)\left|\frac{\mathrm{d}\ln{\sigma}}{\mathrm{d}\ln{M}}\right|\,, &&
\end{flalign}
where $f(\nu)$ is the multiplicity function, for which we use the following functional form \cite{Sheth2001}
\begin{flalign}
 f(\nu) = A\sqrt{\frac{2a}{\pi}}\left[1+(a\nu)^{-p}\right]\nu\exp{\left(-\frac{1}{2}a\nu^2\right)}\,, &&
\end{flalign}
where $\nu=\delta_{\rm c}/\sigma$, with $\sigma$ the variance of the linear matter power spectrum and $A=0.3222$, $a=0.75$ and $p=0.3$ as best fit parameters of the theoretical multiplicity function compared to the one obtained numerically \cite{Sheth2001}.

The mass variance of the smoothed density field is given by
\begin{flalign}
 \sigma^2(M,z) = \frac{1}{2\pi^2}\,\int_{0}^{\infty}\,k^2\,P(k,z)\,W(kR)\,\mathrm{d}k\,, &&
\end{flalign}
where the window function $W(kR) = 3[\sin{(kR)} - (kR)\cos{(kR)}]/(kR)^3$ is the Fourier transform of the top-hat profile and the radius is associated to the halo mass via the relation $M = 4\pi\bar{\rho}_{\rm m,0}R^3/3$ and $\bar{\rho}_{\rm m,0}$ today's average matter density.

We obtain the linear matter power spectrum from a suitably modified version of the Einstein-Boltzmann code \texttt{class}\footnote{\href{https://github.com/lesgourg/class_public/}{https://github.com/lesgourg/class\_public/}} \cite{class}. For internal consistency, the code always evolves the perturbations of every species, therefore, to distinguish between smooth and clustering dark energy, we set \texttt{cs2\_fld}, the variable representing $c_{\rm s}^2$, to $1$ and $0$, respectively\footnote{For simplicity, we perform calculations in the Newtonian gauge.}.

\begin{figure*}[!ht]
 \centering
 \includegraphics[scale=0.35]{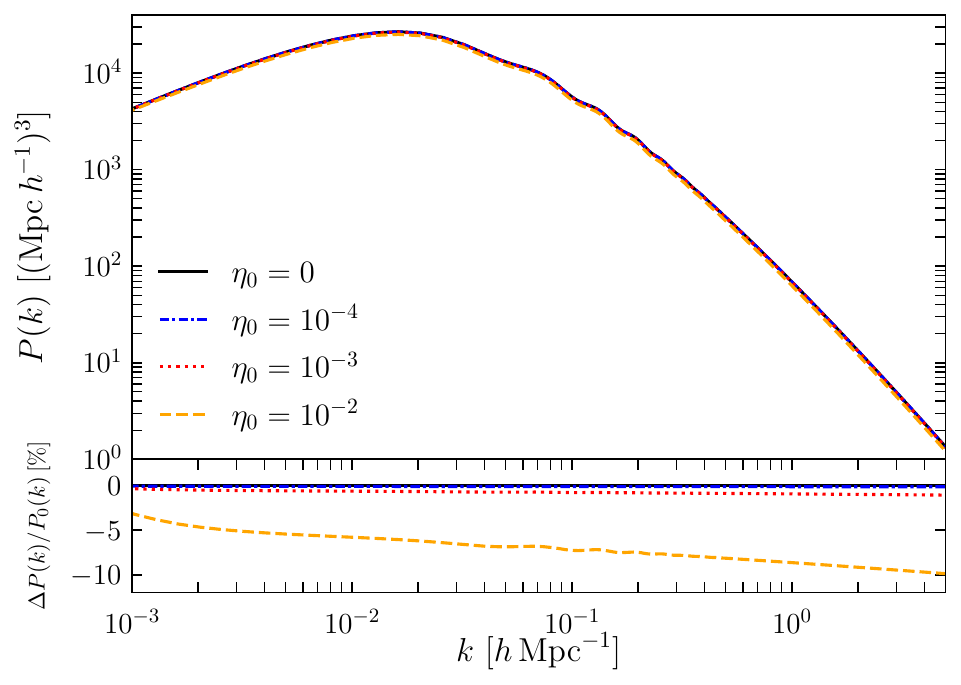}
 \includegraphics[scale=0.35]{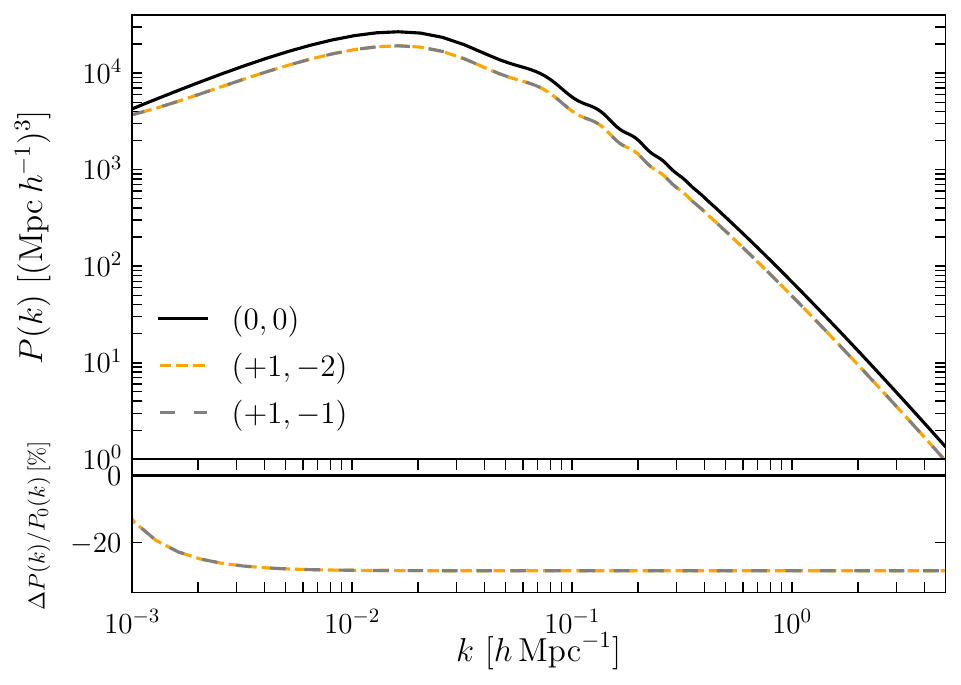}
 \caption{Evolution of the matter power spectrum $P(k)$ at $z=0$ for the $\Lambda$CDM model. In the left (right) panel we show results for dynamical (dissipative pressure) friction. Line styles and colours are as in Fig.~\ref{fig:gf_LCDM}.}
 \label{fig:Pk_LCDM}
\end{figure*}

Before showing our results on the halo mass function, we present in Fig.~\ref{fig:Pk_LCDM} the matter power spectrum for the $\Lambda$CDM model at $z=0$ only. 

Dynamical friction is negligible on all the scales considered, except for the case with the largest $\eta_0$ and on very small scales. In this case, the differences range between $5\%$ and $10\%$. Dissipative friction, instead, as already discussed, shows stronger effects, almost constant along the scales shown and with differences of about 25\% with respect to the case without friction. As expected, differences point towards a lack of power, sign of a slowing down in the evolution of dark matter perturbations. This is due to the fact that we considered that all models have the same initial conditions, in particular the primordial amplitude of perturbations $A_{\rm s}$. This translates into a lower value of the normalization of the matter power spectrum today, $\sigma_8$.

For dark energy models, we find qualitative and quantitative similar results; therefore, we will not show additional plots. Dynamical friction has a very small effect, whereas dissipative pressure friction is substantially more relevant. The exact quantitative effect depends on the particular dark energy model, but its behavior is totally analogous to what is seen for the growth factor, presented in Fig.~\ref{fig:gf_wCDM}. As perturbations in dark energy models are very small, even when $c_{\rm s}^2=0$, we do not see appreciable differences between smooth and clustering dark energy models.

\begin{figure*}[!ht]
 \centering
 \includegraphics[scale=0.35]{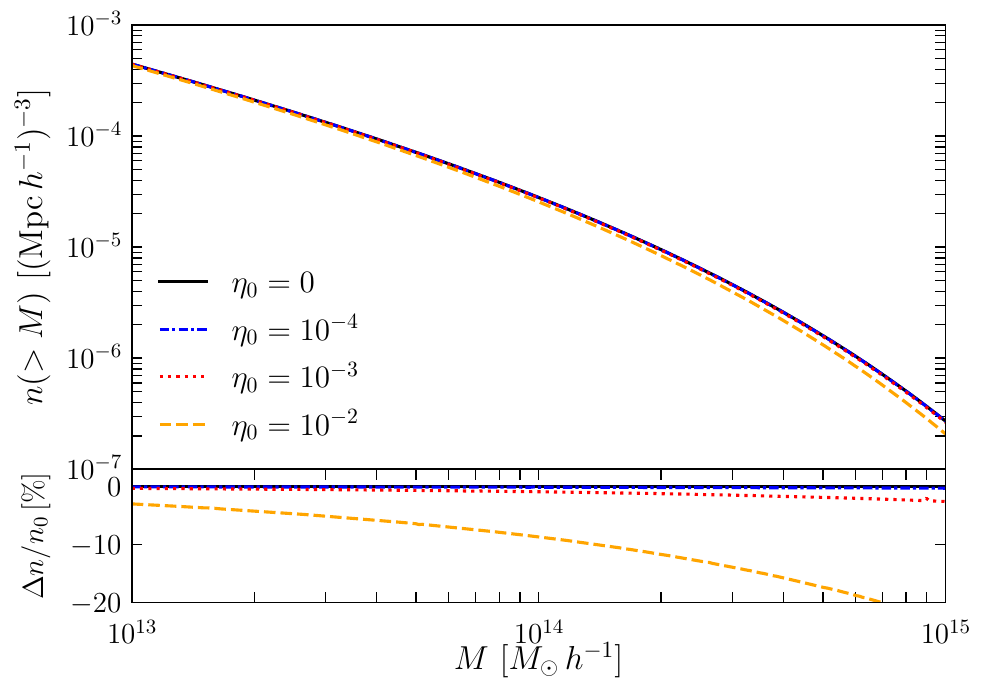}
 \includegraphics[scale=0.35]{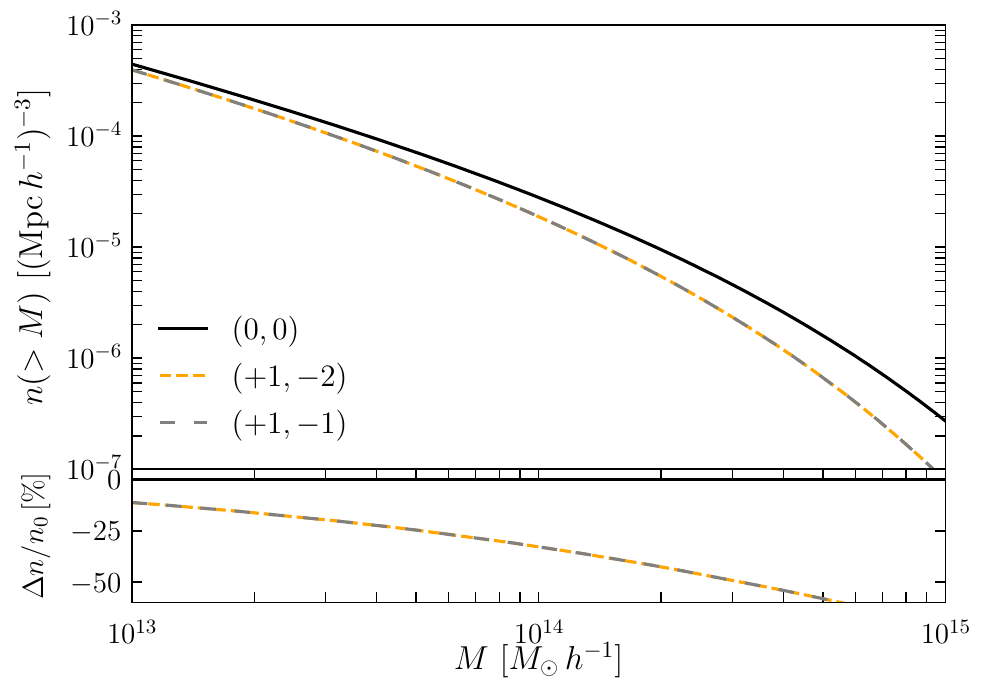}
 \caption{Number of objects above a given mass $M$ for the $\Lambda$CDM model at $z=0$. The left panel shows the effects of dynamical friction while the right panel refers to dissipative pressure friction. Line styles and colours are as in Fig.~\ref{fig:gf_LCDM}.}
 \label{fig:mf_LCDM}
\end{figure*}

\begin{figure*}[!ht]
 \centering
 \includegraphics[scale=0.35]{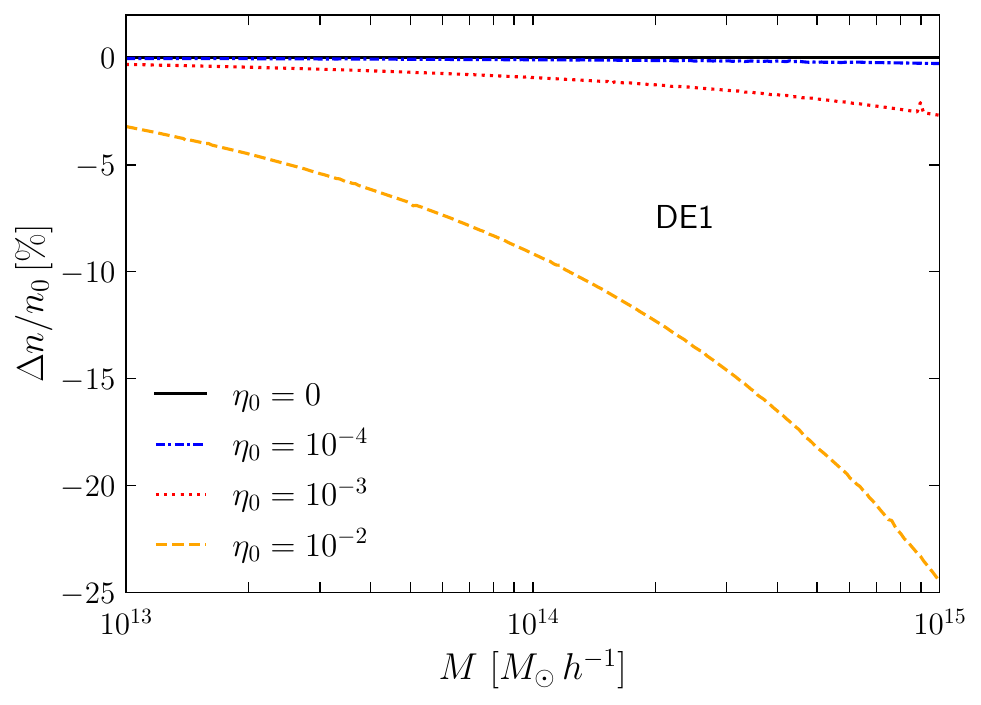}
 \includegraphics[scale=0.35]{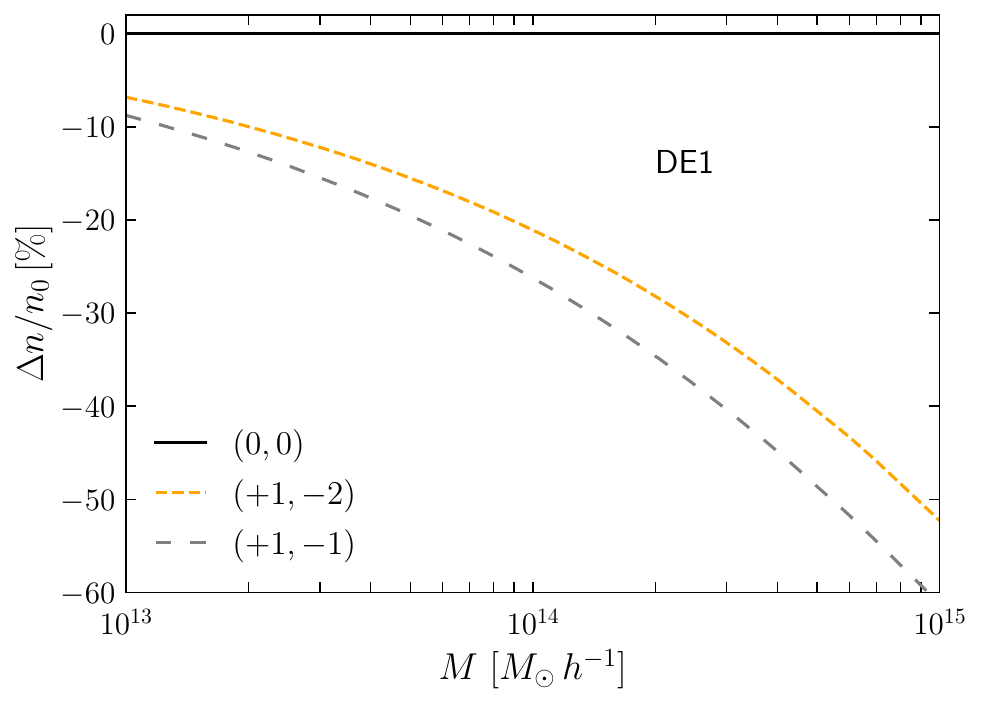}
 \caption{Number of objects above a given mass threshold $M$ for the dark energy model DE1 with respect to the $\Lambda$CDM prediction. We considered dynamical friction (left panel) and dissipative pressure friction (right panel). Line styles and colours are as in Fig.~\ref{fig:gf_LCDM}.}
 \label{fig:mf_wCDM_DE1}
\end{figure*}

We can now discuss the effects of friction on the halo mass function. We present our results for the number of halos above a given mass for the $\Lambda$CDM model in Fig.~\ref{fig:mf_LCDM} and for the smooth dark energy model DE1 in Figs.~\ref{fig:mf_wCDM_DE1}.

Dynamical friction has a small effect on the overall mass function and becomes appreciable only for the most extreme case with $\eta_0=10^{-2}$. In this case, for masses of the order of $7\times 10^{14}\,M_{\odot}\,h^{-1}$, we notice a decrease of about 20\% in the number of objects with $M>10^{15}\,h^{-1}\,M_{\odot}$ for $\eta_0=0.01$, but also a decrease of roughly 10\% for objects with masses similar to galaxy groups. For lower values of $\eta_0$, differences with respect to $\Lambda$CDM are much smaller and probably difficult to measure. These values are approximately the same for all six models. On the same scales, the effect of dissipative pressure friction is about three times higher. As we evaluated the halo mass function today, this is a consequence of the lack of power highlighted in the matter power spectrum. We also remind the reader that for the $\Lambda$CDM model, the effect of $\Gamma$ is irrelevant. This result strengthens the previous discussion on the possibility of placing tight constraints on these models.

For the smooth case, all the models have the same qualitative behavior when considering dynamical friction and the differences with respect to a frictionless model are largely independent of the particular equation of state used. As expected, differences increase at higher masses and for all the cases here considered we observe a decrease in the number of objects which form. This is consistent with the fact that friction dampens perturbations. The higher $\eta_0$, the larger the differences. For $\eta_0=10^{-4}$, deviations from the standard case where dynamical friction is absent are again negligible. The differences become appreciable for $\eta_0=10^{-3}$ and reach a value of about 2\%--3\% for masses of the order of $10^{15}\,M_{\odot}\,h^{-1}$, to reach 20\%--25\% for $\eta_0=10^{-2}$.

More interesting is the case of dissipative pressure friction. The differences are enhanced with respect to the case with $\eta_2 = 0$ and can reach up to 75\% fewer objects. A particular case is model DE6, whose equation of state is much smaller than $-1$. For this model, the suppression of structures is substantial, of the order of $50\%$ already for objects above $10^{13}\,M_{\odot}\,h^{-1}$.

The analysis shows another interesting point. Whilst the behavior of dynamical friction is coherent among the models, and percentage differences are roughly the same, for dissipative pressure friction this is not the case any more. First of all, the exact difference strongly depends on the background equation of state. In fact, excluding model DE6, we observe that differences range from 50\% to 75\% for masses of $10^{15}\,M_{\odot}\,h^{-1}$. But more importantly, analogously to what happens for the growth factor, we see that the two cases with $\Gamma=-1$ and $\Gamma = -2$ have a different behavior with respect to each other: in fact, for models with quintessence behavior (models DE1 and DE3), the model with $\Gamma=-2$ predicts more objects than when $\Gamma=-1$. Opposite situation where the model is always, or at least for some time, in the phantom regime. Thus, the lower the equation of state, the stronger the differences.

\section{Outlooks}\label{sect:conclusions}

In this paper, we investigate the effects of friction on large-scale structures. In this respect, we focussed on both smooth and clustering dark energy models and limited our analysis to general relativity and to models with a linearly evolving equation of state, as these are more in line with observational data.

Specifically, to describe the role of friction, we examined two main backgrounds. The first considers friction proportional to the Hubble rate, while the second involves a more complex correction primarily related to the dark energy pressure, \textit{i.e.}, involving relativistic effects due to the pressure of dark energy.

We analyze the effects of friction on both linear and nonlinear perturbation regimes employing the foundational principles of spherical collapse theory. To do so, we studied the evolution of perturbations, and we determined the appropriate initial conditions to develop such perturbations.

Our findings revealed that the most significant discrepancies occur in the second hierarchical model, as it severely affects both linear and nonlinear perturbations. In this respect, we detailed the effects and consequences of friction on large-scale structures, highlighting the key differences from standard cases that do not incorporate such effects from the outset. 

In addition, we developed the consequences of the modified evolution of the cumulative mass function in the two aforementioned schemes. Consequences and discussions towards the underlying effects are reported, showing that our expectations are in line with current comprehension of our data. We demonstrated that the first recipe, dynamical friction proportional to the Hubble function, affects the mass function by 5\% at most, while for the second model differences can be up to 75\% for the range of masses invesigated in this work.

We also considered how perturbations of dark energy affect the halo mass function. It turns out that the effects of clustering of dark energy are significantly more important than friction and that the contributions of perturbations are strongly dependent on the background equation of state.

Consequently, with the severe exception of the second background treatment, our results seem to suggest that there is no \emph{a priori} reason to exclude the presence of friction, indicating that dissipation can potentially have a significant limiting effect on clustering.

In future work, we intend to extend our effort by involving more complicated friction models and also investigating the consequences on structure formation. Additional studies will be performed to show how dark energy clustering can also be influenced by dark energy, and/or we will clarify whether a possible interaction between the dark sector and friction may occur.

\section*{CRediT authorship contribution statement}
\textbf{Francesco Pace}: Supervision, Project administration. \textbf{Orlando Luongo}: Writing – review \& editing, Writing – original draft, Investigation. \textbf{Antonino Del Popolo}: Validation, Conceptualization.

\section*{Declaration of competing interest}
The authors declare that they have no known competing financial interests or personal relationships that could have appeared to influence the work reported in this paper.

\section*{Acknowledgements}
\noindent
F.P. acknowledges partial support from the INFN grant InDark and from the Italian Ministry of University and Research (\textsc{mur}), PRIN 2022 `EXSKALIBUR – Euclid-Cross-SKA: Likelihood Inference Building for Universe's Research', Grant No.\ 20222BBYB9, CUP C53D2300131 0006, and from the European Union -- Next Generation EU. FP also acknowledges support from the FCT project ``BEYLA -- BEYond LAmbda" with ref. number PTDC/FIS-AST/0054/2021.

\appendix

\section{Dark energy}\label{sect:DE}
In the next two subsections we will present results for the dark energy models considered in this work. We consider both smooth and clustering dark energy, but we will show figures about smooth dark energy models only, as the effect of clustering dark energy is very small, and qualitatively we obtain the same results in both cases.

\begin{figure*}
 \centering
 \includegraphics[scale=0.4]{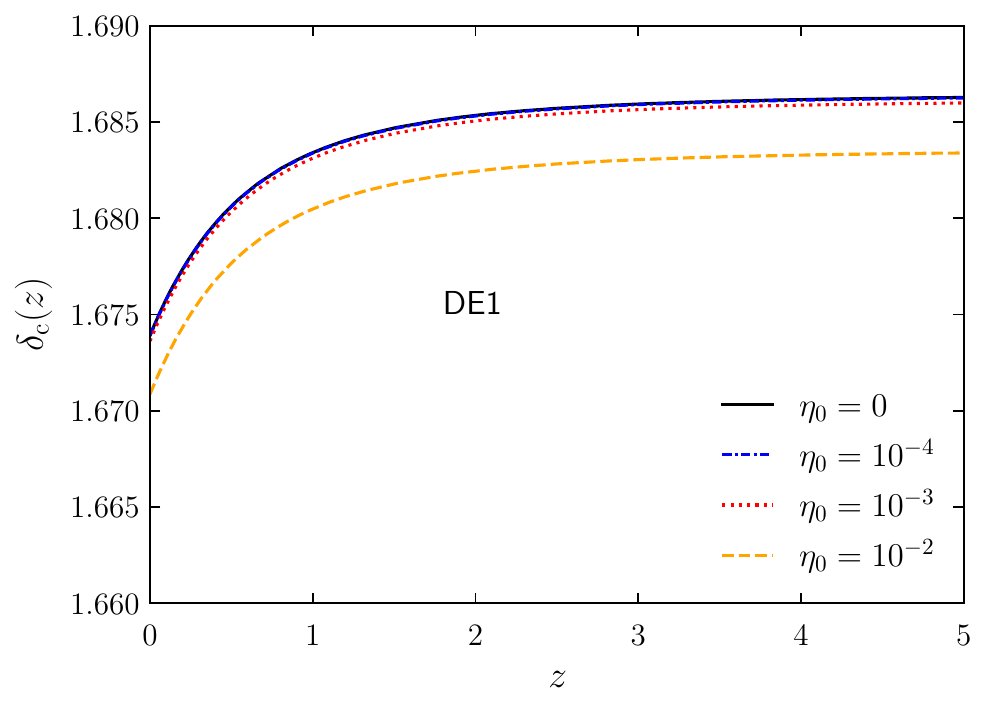}
 \includegraphics[scale=0.4]{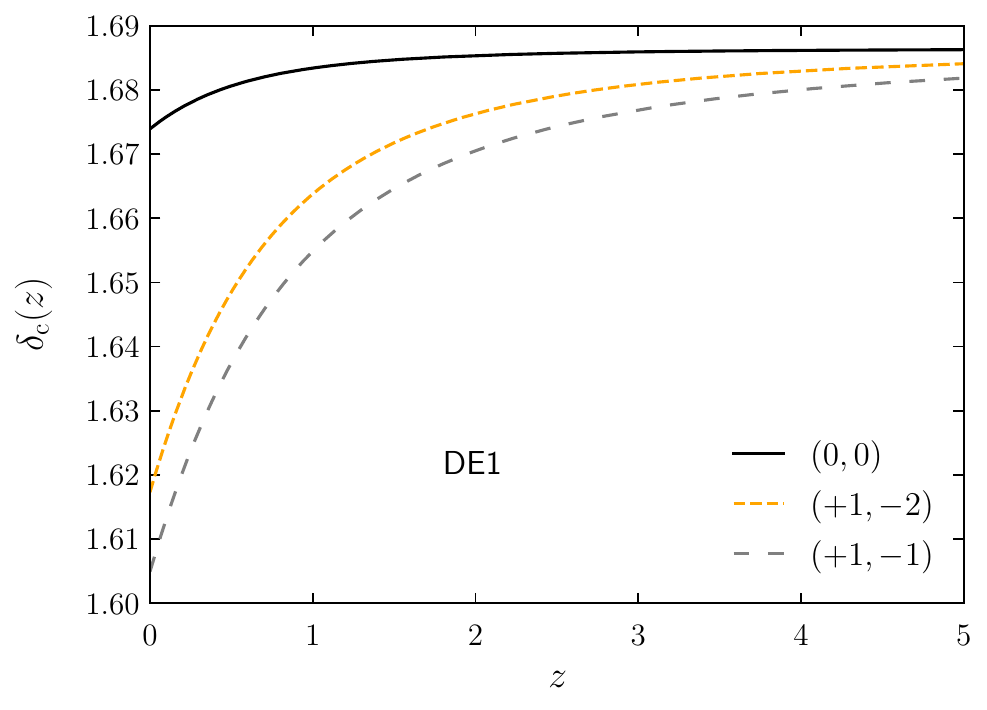}
 \caption{Evolution of the linearly extrapolated overdensity $\delta_{\rm c}$ for the model DE1 for dynamical friction (left panel) and dissipative pressure friction (right panel). Line styles and colours are as in Fig.~\ref{fig:gf_LCDM}.}
 \label{fig:deltaC_wCDM_DE1}
\end{figure*}

\subsection{Smooth dark energy}
For smooth dark energy, we can infer conclusions analogous to the $\Lambda$CDM model. The effects of dynamical friction on the linearly extrapolated overdensity parameter $\delta_{\rm c}$ (left panel of Fig.~\ref{fig:deltaC_wCDM_DE1}) are small and of the same order of magnitude as the $\Lambda$CDM model and deviations from the case with $\eta_0 = 0$ are largely independent of the particular model considered.

For dissipative pressure friction (right panel of Fig.~\ref{fig:deltaC_wCDM_DE1}), as for the $\Lambda$CDM model, the overall amplitude of $\delta_{\rm c}$ is lower than when friction is not considered. The strength of the effect depends on the particular dark energy model. Models DE2, DE5 and DE6 show the strongest deviations, while for the other three models the differences are less pronounced. Unlike before, now the particular value of $\Gamma$ affects the final behavior of the model. In fact, for the phantom model DE6, when $\Gamma=-2$, the effect of friction is so large that the initial overdensity must reach unrealistic values so that the perturbations can collapse. For this reason, we do not show the results obtained. This is the same behavior we observed when studying the growth factor.

\begin{figure*}
 \centering
 \includegraphics[scale=0.4]{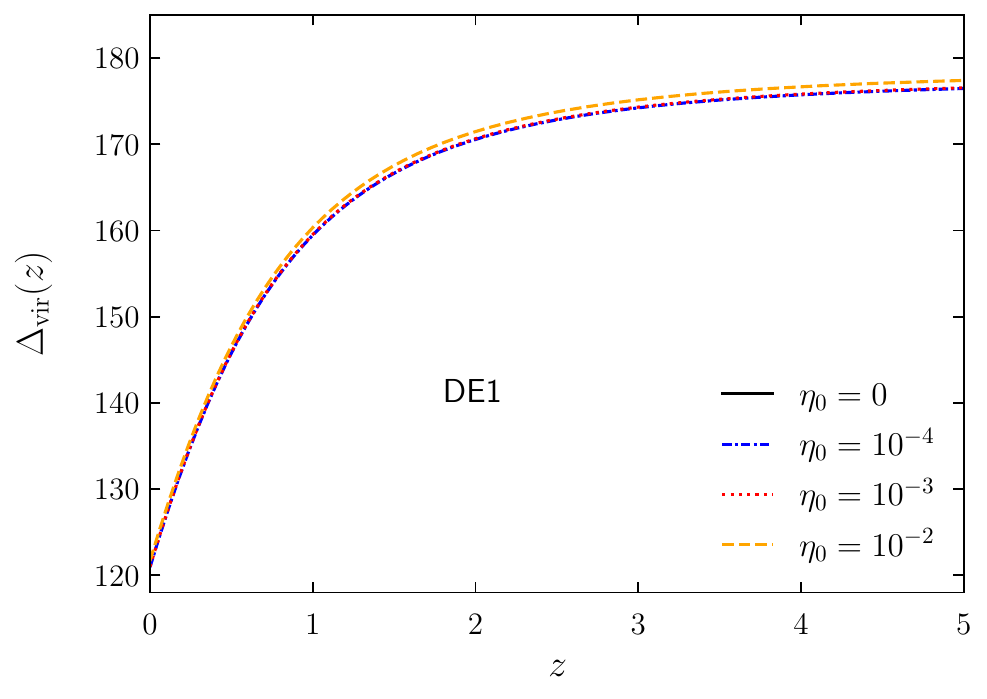}
 \includegraphics[scale=0.4]{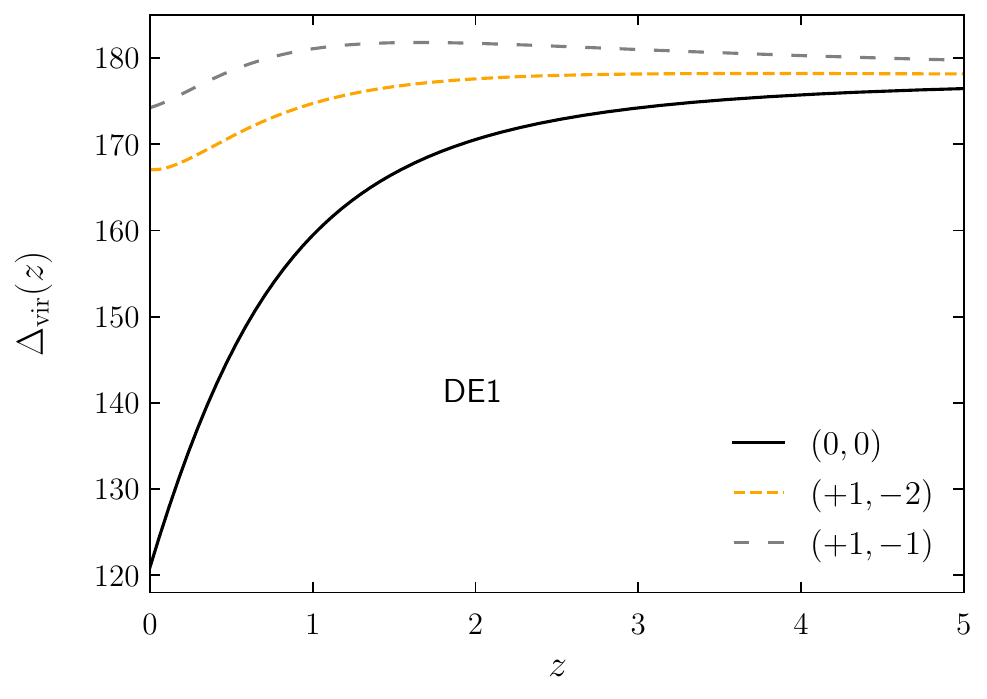}
 \caption{Evolution of the virial overdensity $\Delta_{\rm vir}$ for the model DE1 for dynamical friction (left panel) and dissipative pressure friction (right panel). Line styles and colours are as in Fig.~\ref{fig:gf_LCDM}.}
 \label{fig:deltaV_wCDM_DE1}
\end{figure*}

Regarding the effects of friction on the evolution of virial overdensity $\Delta_{\rm vir}$, the results obtained are in line with what found for $\Lambda$CDM. In the left (right) panel of Fig.~\ref{fig:deltaV_wCDM_DE1} we present our results for dynamical friction (dissipative pressure friction). Also for dark energy models, dynamical friction has a very modest impact with differences of the same order of magnitude found for the $\Lambda$CDM model. Dissipative pressure friction obviously has a much stronger effect, and we see an increase in the value of the virial overdensity. For models DE2 and DE5, in fact, the virial overdensity first increases and then decreases. It is also interesting to see that for model DE3, the virial overdensity is almost constant over all the redshifts investigated.

\subsection{Clustering dark energy}

In this section, we discuss results for clustering dark energy models, where we allow for perturbations of dark energy. Their strength is related to the value of the effective sound speed, defined, as previously discussed, by $c^2_{\rm eff} = \delta P_{\rm de}/\delta\rho_{\rm de}$. To fix it, one would need to consider a scalar field and its Lagrangian. We will therefore consider it as a free parameter. Quintessence models have $c^2_{\rm eff} = 1$ and their perturbations are negligible on the scales of interest \cite{Unnikrishnan2008}. For $k$-essence models we can achieve smaller values and in this work we will consider $c^2_{\rm eff} = 0$ for which dark energy perturbations are larger and the dark energy fluctuations behave like matter.

Once again, we will consider the linear (represented by the growth factor) and the nonlinear evolution (represented by $\delta_{\rm c}$ and $\Delta_{\rm vir}$), but we will not present any figure, as there are no interesting nor special features even in this case.

In the linear regime, dark energy perturbations are always very subdominant with respect to matter perturbations, and the evolution of $\delta_{\rm m}$ is totally unaffected by them. Similar results are obtained when considering the evolution of the linearly extrapolated overdensity $\delta_{\rm c}$ and of the virial overdensity $\Delta_{\rm vir}$. We did not find appreciable differences with respect to the smooth case.

When dark energy is clustering, there is always the question of what is the correct definition of the density perturbation to be used. So far we have assumed that it is enough to consider dark energy affecting matter perturbations via modifications of the Poisson equation, and we found that dark energy perturbations do not affect matter perturbations appreciably. One could also assume that what really matters is not matter density perturbations, but rather the total density perturbation, defined as:
\begin{flalign}
 \delta(a) = \delta_{\rm m}(a) + \frac{\Omega_{\rm de}(a)}{\Omega_{\rm m}(a)}\delta_{\rm de}(a)\,, &&
\end{flalign}
where it is understood that perturbations are either linear or nonlinear according to the particular quantity investigated. Here, we will limit our analysis to the evolution of linear perturbations, in particular of $\delta_{\rm c,tot}$.

\begin{figure*}
 \centering
 \includegraphics[scale=0.4]{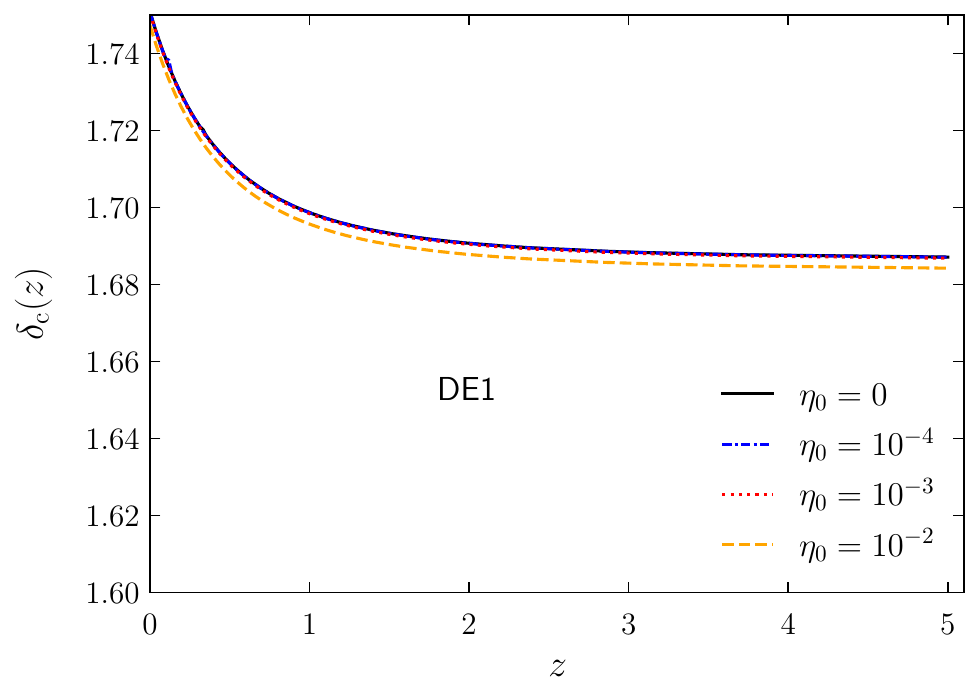}
 \includegraphics[scale=0.4]{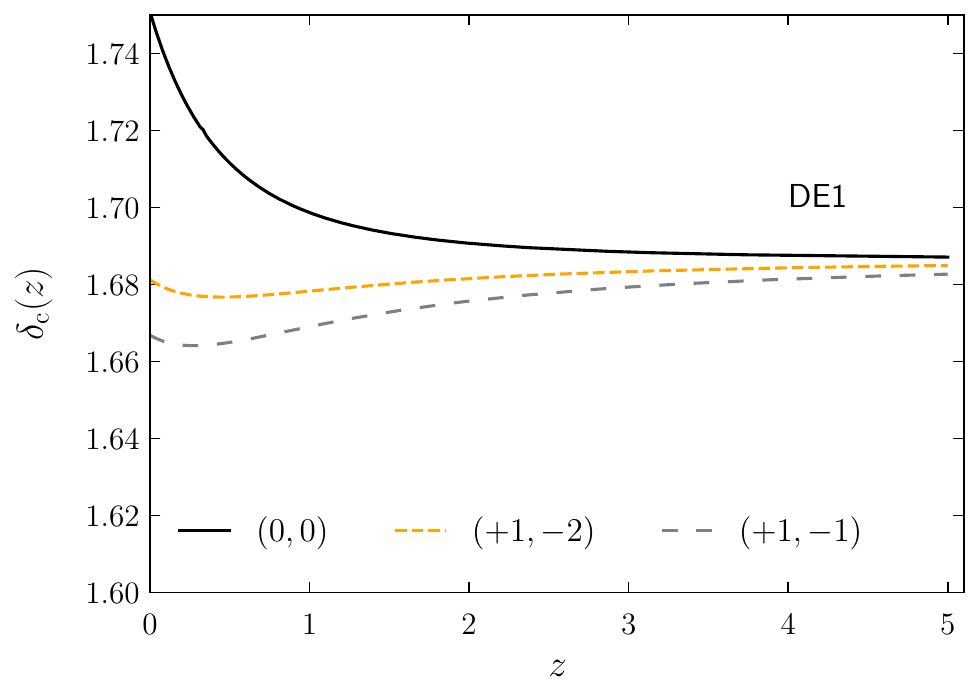}
 \caption{Evolution of the total linearly extrapolated overdensity $\delta_{\rm c}$ for the model DE1 for dynamical friction (left panel) and dissipative pressure friction (right panel). Line styles and colours are as in Fig.~\ref{fig:gf_LCDM}.}
 \label{fig:deltaC_Tot_wCDMc_DE1}
\end{figure*}

Our results are presented in Fig.~\ref{fig:deltaC_Tot_wCDMc_DE1}. We can appreciate that the behavior of $\delta_{\rm c, tot}$ is rather different from before and strongly depends on the particular parameterization of the equation of state used. Considering the effect of dynamical friction, we observe for models DE2, DE5 and DE6 the typical behavior obtained before. The linearly extrapolated overdensity approaches the EdS value (dark energy perturbations are negligible and furthermore scaled by $\Omega_{\rm de}$) and decreases at late times. For the other models, instead, we recover the EdS value at an early time but at late times $\delta_{\rm c, tot}$ grows, showing that dark energy perturbations can have an appreciable effect. It is also interesting to note that the effective value of dynamical friction becomes rather irrelevant, as all the models behave similarly to the case of no friction with $\eta_0 = 0$.

More complex is the analysis for dissipative pressure friction. The results depend not only on the equation of state used, but also on the particular value of the parameter $\Gamma$. Thus, we observe a similar trend to that found in Fig.~\ref{fig:deltaC_Tot_wCDMc_DE1}. Models DE2, DE5 and DE6 behave as one would expect for smooth dark energy or for associating $\delta_{\rm c}$ to matter only. The other three models, instead, have EdS-like behavior at early times and grow, rather than decrease, at late times. It is also interesting to see that the dependence on the exponent $\Gamma$ becomes very shallow.

This discussion highlights an important point about the correct definition of density perturbations, as although $\delta_{\rm de}$ is smaller than $\delta_{\rm m}$ in general and has limited effects on the evolution of the latter, it can substantially affect our observables if we use $\delta_{\rm tot}$ instead of $\delta_{\rm m}$, as was also demonstrated in discussing virialization in \cite{Pace2022}.

\subsection{Corrections to the halo mass function}
When dark energy clusters, the halo mass function is also modified. To take into account these modifications, we follow the work of \cite{Creminelli2010}. According to the authors, one can scale the halo mass function calculated for a smooth model with the Sheth-Tormen \cite{Sheth2001} formulation with the halo mass function calculated with the Press-Schechter \cite{Press1974} formalism.

\begin{figure*}
 \centering
 \includegraphics[scale=0.4]{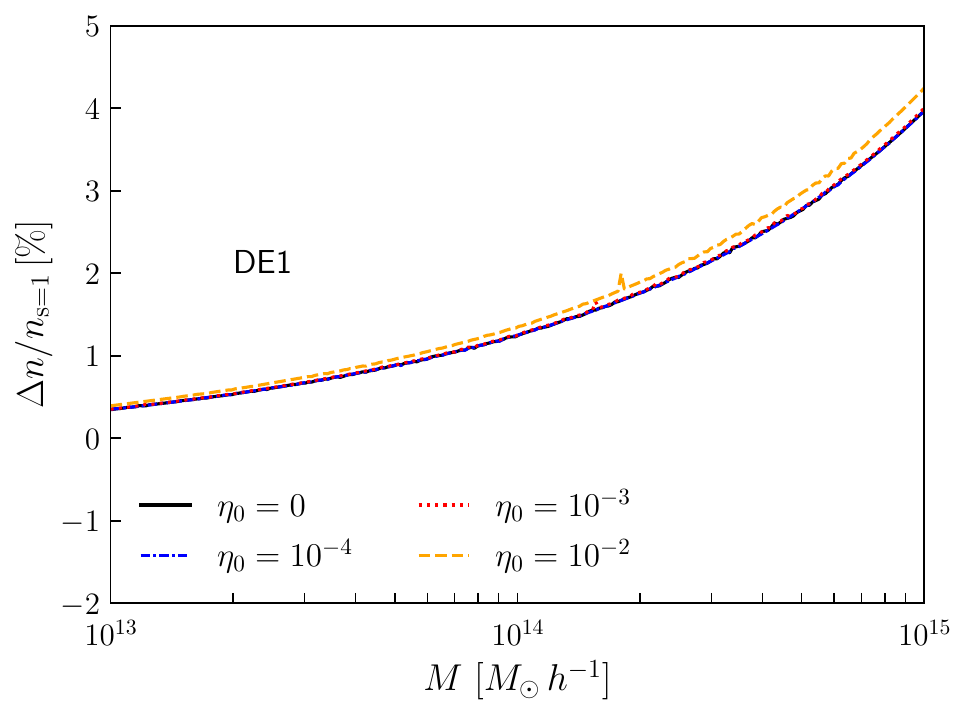}
 \includegraphics[scale=0.4]{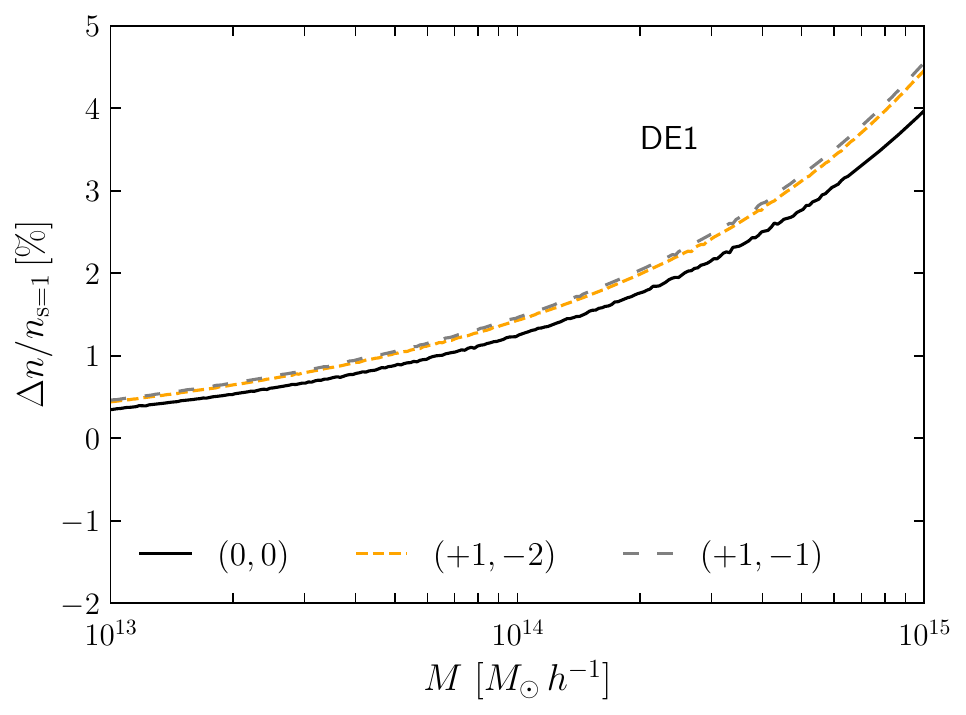}
 \caption{Number of objects above a given mass threshold $M$ for the model DE1, when dark energy is perturbed. We considered dynamical friction. Left panel shows results for dynamical friction, while the right panel for dissipative pressure friction. Line styles and colours are as in Fig.~\ref{fig:gf_LCDM}.}
 \label{fig:mf_wCDMc_DE1}
\end{figure*}

The halo mass function for smooth models is thus \cite{Creminelli2010}
\begin{flalign}\label{eq:ps_scaling}
    \frac{\mathrm{d}n}{\mathrm{d}\ln{M}} = \frac{\mathrm{d}n_{\rm{ST, s=1}}}{\mathrm{d}\ln{M}} \frac{\mathrm{d}n_{\rm{PS, s=0}}/\mathrm{d}\ln{M}}{\mathrm{d}n_{\rm{PS, s=1}}/\mathrm{d}\ln{M}}\,, &&
\end{flalign}
where it is understood that when the particular dark energy model is clustering, one needs to use the appropriate matter power spectrum and linear overdensity parameter $\delta_{\rm c}$. The reason for this choice is that the ratio of the Press-Schechter functions will mitigate its shortcomings. We present our results in Fig.~\ref{fig:mf_wCDMc_DE1}, where we consider the ratio between the clustering model and the corresponding smooth model.

For dynamical friction, we notice that the dominant contribution is that of clustering. Whether we predict more or less objects depends on the model considered. If the model has a quintessence equation of state, then clustering predicts more objects, with differences ranging from the subpercent level for models DE1 and DE5 to a couple of percent for model DE3 for masses of the order $10^{13}\,M_{\odot}\,h^{-1}$. At higher masses, the differences can be greater than 10\%. We notice that these substantial differences appear for model DE3, whose equation of state is the most distant from that of the $\Lambda$CDM model.
For phantom models, we observe a decrease in structure formation, generally of a few percent and more substantial for model DE6, as it is the model more different from $\Lambda$CDM.

Very similar results are obtained, from both a qualitative and a quantitative point of view, for dissipative pressure friction. Also in this case, clustering is the dominating effect and differences are maximized for models DE3 and DE6. Except for the latter, models with different values of $\Gamma$ behave very similarly to each other.

We also notice that our results are qualitatively in good agreement with the original work of \cite{Creminelli2010}.

\section*{Data availability}
No data was used for the research described in the article.

\bibliography{main.bbl}

\end{document}